\documentclass[aps,prl,preprint,preprintnumbers,amsmath,amssymb,floatfix]{revtex4-2}

\usepackage[bookmarksnumbered,pdfpagelabels=true,plainpages=false,colorlinks=true,linkcolor=blue,citecolor=blue,urlcolor=blue]{hyperref}
\usepackage{graphicx}% Include figure files
\usepackage{dcolumn}% Align table columns on decimal point
\usepackage{bm}% bold math
\usepackage{morefloats}
\usepackage[bookmarksnumbered,pdfpagelabels=true,plainpages=false,colorlinks=true,linkcolor=blue,citecolor=blue,urlcolor=blue]{hyperref}
\usepackage{bbm}
\usepackage{tabularx}
\usepackage{cancel,soul}
\usepackage{nicefrac, xfrac}
\usepackage{ulem}
\usepackage{bbold}

\usepackage{xcolor}
\usepackage{nicefrac, xfrac}
\def \usach {Departamento de F\'isica, Universidad de Santiago de Chile, 9170124, Santiago, Chile.}
\def \cedenna {Centro  de Nanociencia y Nanotecnología CEDENNA, Avda. Ecuador 3493, Santiago, Chile.}
\def \fcfm {Departamento de F\'isica, FCFM, Universidad de Chile, Santiago, Chile.}
\def \ua {Grupo de Investigación en Física Aplicada (GIFA), Facultad de Ingenier\'ia, Universidad Aut\'onoma de Chile, Av. Pedro de Valdivia 425, Providencia, Santiago, Chile.}
\def \ufv {Departamento de Física, Universidade Federal de Viçosa, Av. PH Rolfs s/n, 36570-900, Viçosa, Brazil.}

\begin{document}

%\preprint{APS/123-QED}
\title{Bimerons as Edge states in Thin Magnetic Strips}
%\title{Edge States of Bimerons in Magnetic strips}

%\\{\color{blue}Bimeron racetrack without Hall effect}\\{\color{red}Engineering bimeron racetrack}\\{Overcoming bimeron Hall effect in racetracks}}% Force line breaks with \\
%\thanks{A footnote to the article title}%

\author{Mario Castro$^{1}$}
\email{mario.castrob@usach.cl}
\author{David Gálvez-Poblete$^{2,3}$}
\author{Sebastián Castillo-Sepúlveda${}^{4}$}
\author{Vagson L. Carvalho-Santos${}^{5}$}
\author{Alvaro S. Nunez${}^{1,2}$}
\author{Sebastian Allende${}^{2,3}$}

\affiliation{${}^{1}$\fcfm}
\affiliation{${}^{2}$\cedenna}
\affiliation{${}^{3}$\usach}
\affiliation{${}^{4}$\ua}
\affiliation{${}^{5}$\ufv}

\begin{abstract}
Magnetic bimerons are potential information carriers in spintronic devices. Bimerons, topologically equivalent to skyrmions, manifest in chiral magnetic systems with in-plane magnetization due to anisotropies or external magnetic fields. Applications demanding their current-driven motion face significant challenges, notably the bimeron Hall effect, which causes transverse movement and annihilation at nanomagnet borders.
This study addresses the problem of stabilizing bimeron propagation under current-driven conditions. We demonstrate that bimerons can propagate through thin ferromagnetic strips without annihilation when the easy-axis anisotropy and the electric current are orthogonal. Our findings indicate that below a threshold value of current, the repulsion between the bimeron and the strip boundary allows for stable soliton propagation, even in bent regions. This phenomenon extends to bimeron chains, which propagate parallel to the current flow. By enabling stable long-distance propagation, our results open new avenues for developing bimeron-based racetrack memory devices, enhancing the efficiency and reliability of future spintronic applications.
\end{abstract}

\maketitle

%\tableofcontents

\noindent{\textbf{\textit{Introduction -}} } 
Spintronics is an exciting and rapidly advancing field in condensed matter physics, with the potential to improve transmission, processing, and data storage devices\cite{Yang2021, Ref1, Ref2, Ref3, He2021}. A key attractive idea in spintronics technology involves the racetrack memory device \cite{Parkin-Sci, Parkin-Nat,JSampaio2013, Blasing2020, Parkin2022, Gu2022, FernndezPacheco2022}. In this conceptual framework, magnetic textures, which propagate along a strip under the influence of spin-polarized currents, function as discrete bits for data storage \cite{Parkin-Sci, Gu2022}. A diverse array of magnetization configurations is poised to revolutionize racetrack memory technology, serving as robust information carriers. These include dynamic domain walls \cite{Parkin-Sci, Gu2022,Indian-IEEE}, vortices \cite{Geng-JMMM}, and ferro- and antiferromagnetic skyrmions \cite{JSampaio2013, Tomasello-Sci, Tomasello-JPD, Liang-PRB}. In addition, complex structures such as skyrmionium \cite{Zhang-PRB, Cotrina-APL}, hopfion \cite{Liu-PRL, Wang-PRL}, and bimeron \cite{Kim-PRB, CastroMA2023, Li-NPJ, Liang-PRB2} are also being considered, each offering unique advantages and possibilities in the realm of advanced data storage solutions.

Despite its promising potential, the practical implementation of the racetrack memory concept encounters significant challenges in achieving precise control over the transport of magnetization collective modes along the strips. Specifically, in the domain walls, their propagation driven by electric currents is restricted by the so-called Walker limit \cite{Walker}. This limitation manifests as a reduction in velocity, attributed to the complex interplay of upward and backward motions that arise due to the dynamic energy changes occurring during their propagation \cite{Mougin, Guslienko}. Domain walls not only exhibit direction-dependent dynamical behavior \cite{Parkin-Nat3, Gu2022}, but also are subject to pining effects due to defects \cite{Yuan-PRB, Beach-Nat} and curved regions \cite{Yershov-PRB, Bittencourt-PRB} of the strip. The electric current required to mobilize vortices and skyrmions along the nanotrack at least two orders of magnitude smaller than that needed for domain wall motion \cite{JSampaio2013, Geng-JMMM}. However, the emergence of the vortex and skyrmion Hall effect \cite{Nagaosa-Nat, Lizius-Nat1, Lizius-Nat2} introduces a dynamic component to the magnetic texture velocity perpendicular to the current direction. This phenomenon leads to the annihilation of skyrmions and vortices at the borders of the nanotrack, which presents a challenge to the stability and lifetime of these magnetic textures. The skyrmion Hall effect can be avoided if a skyrmionium nucleates in the magnetic system \cite{Zhang-PRB}. However, these quasiparticles have low stability against thermal fluctuations, with a lifetime on the order of picoseconds at room temperature \cite{Jiang-PRR}, and their experimental nucleation is still challenging \cite{Yang-Nat}. Current experimental techniques have enhanced the quality of fabricated nanoparticles, allowing the nucleation and control of three-dimensional solitons, such as hopfion \cite{Sutclife-PRL, Sutclife, Brataas}. The analysis of hopfion motion under electric currents revealed a complex dynamical behavior in which breathing and rotation of the magnetic texture avoid proper control of its propagation along the magnetic body \cite{Liu-PRL, Wang-PRL, Nagaosa-PRL}.

Theoretical propositions have been presented to solve the challenges of running racetrack devices. The main ones are about avoiding the skyrmion Hall effect through different techniques, such as including extra anisotropy \cite{Toscano2020,Zhang-Nanoscale,Guo2022} or geometrical constraints \cite{Purnama,Jin-NPJ} along the nanotrack borders. However, the experimental implementation of these propositions is not yet available. A promising solitonic texture that can be used as an information carrier in racetrack devices is the bimeron \cite{Gobel-PRB,Silva-PRB,Yakolev-PRB, Rosales2023, Nagase2021, Ohara2022, Amari2024, Yu2023}, whose experimental observation has been reported in Py film \cite{GaoN2019}, in chiral magnets such as Co-Zn-Mn (110) thin films \cite{NagaseT2021}, and in ferromagnetic multilayers under an external magnetic field\cite{Ohara2022}. Theoretical work predicts that the Hall angle of the bimeron depends on the relative direction between the line that connects their centers and the direction of the electric current \cite{Zarzuela2020} in such a way that bimerons can displace longer distances along the strip before being annihilated at its border \cite{Alane}. Also, including magnetic field gradients compensating the Magnus force can overcome or suppress the skyrmion Hall effect in bimerons \cite{Shen-PRB}. 

 An interesting finding about bimerons is their interaction with edge states in a nanotrack under an external magnetic field, where collinear in-plane spins are induced at low field values, leading to an asymmetric bimeron-edge interaction potential \cite{Leonov}. In this work, we demonstrate that this interaction with the nanotrack edge can induce bimeron confinement, allowing them to propagate along the strip without annihilating at the edges, as long as the orientation of the easy-axis anisotropy and the electric current are orthogonal.
Under these conditions, the bimeron undergoes the Hall effect, yet its destruction is circumvented through its interaction with the strip's border. As the bimeron moves along the edge,  its velocity increases, more than twice the speed observed within the bulk of the strip. The bimeron-edge interaction allows it to traverse the full length of the strip when subjected to an electric current below a threshold value, even in the presence of bent sections. Essentially, the bimeron dynamics establishes an edge state within the strip. We also show that bimeron chains exhibit propagation parallel to the direction of the current. Most aspects of this research focus on describing the bimeron edge state for an anisotropy lying in the plane of the strip. We have also extended our analysis to the scenario where the anisotropy is perpendicular to the plane, whose results are presented in supplemental materials. Our findings blaze the trail for many innovative bimeron-based racetrack memory devices.

\begin{figure}[ht!]
    \centering
    \includegraphics[width=0.9\linewidth]{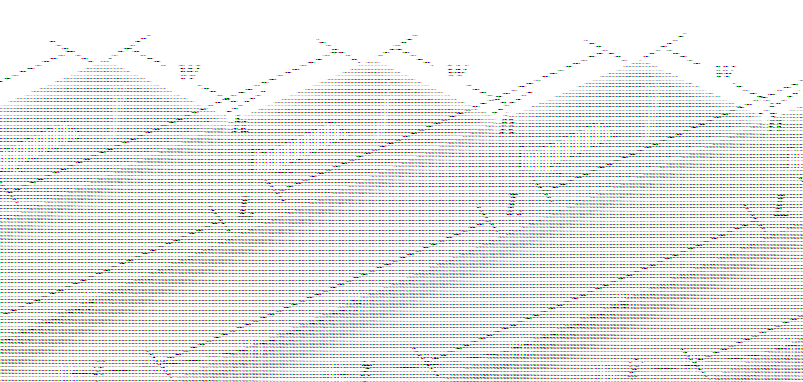}
    \caption{Schematic representation of the geometrical dimensions of a magnetic strip on which a bimeron will move.}
    \label{fig1}
\end{figure}

\noindent{\textbf{\textit{Stabilization of isolated bimeron in chiral magnet with easy-plane anisotropy -}} } We consider bimerons in a thin ferromagnetic (FM) strip deposited on a non-magnetic (NM) substrate. The FM and NM materials interface induces an interfacial Dzyaloshinskii-Moriya (DM) interaction. The strip dimensions are length $L = 3 \ \mu \mathrm{m}$, width $w = 256 \ \mathrm{nm}$, and height $h = 1 \ \mathrm{nm}$, as depicted in Figure \ref{fig1}. The magnetic parameters used are exchange stiffness constant $A_{ex} = 15 \ \mathrm{pJ/m}$, saturation magnetization $M_s = 580 \ \mathrm{kA/m}$, and interfacial Dzyaloshinskii-Moriya constant $D_i = 2.9 \ \mathrm{mJ/m^2}$, typical of a Co/Pt interface \cite{Saha2019}. The easy uniaxial anisotopy axis is defined by $\hat{y}$ direction, where the anisotropy constant is $K=1.4 \times 10^5$ $\mathrm{J/m}^3$.  Due to an asymmetric bimeron-edge interaction potential \cite{Leonov}, the direction of the uniaxial anisotropy axis is the key point of our article. 

%in {\color{red}{The easy-plane anisotropy constant $K$ varies between $1.0 \cdot 10^5$ and $1.8 \cdot 10^5 \ \mathrm{J/m}^3$, with the anisotropy axis defined as $(\cos\theta_K, \sin\theta_K, 0)$. The angle $\theta_K$ ranges from $0^\circ$ to $x^\circ$.}}
%for which a bimeron (o quizas magnetic soliton) has been stabilized in previous works [].

Numerical calculations were performed using the Mumax$^3$ open framework \cite{Mumax3}. The dynamic of reduced magnetization $\vec{m}=\vec{M}/M_s$ is described by the LLG equation, including the term $\vec{\tau}_{SOT}$, which represents the spin-orbit torque (SOT) specific to the system under consideration  (ferromagnetic/heavy-metal layers) \cite{Tomasello2014,Bttner2017}. This equation is given by:

\begin{equation}\label{eq:1}
    \frac{\partial \vec{m}}{\partial t}=-\gamma \vec{m} \times \vec{\rm H}_{\mathrm{eff}}+\alpha \vec{m}\times  \frac{\partial \vec{m}}{\partial t}+\vec{\tau}_{SOT}.
\end{equation}

\noindent Here, $\gamma$ and $\alpha$ are the gyromagnetic ratio and the Gilbert damping constant, respectively. $\vec{\rm H}_{\rm eff}=-(1/\mu_0 M_s)(\delta \omega_E/\delta \vec{m})$ is the effective magnetic field, with $\omega_E$ the energy density, determined by exchange, magnetostatic, DM and anisotropy contributions.  The SOT term is given by

\begin{align}\label{eq:2}
   &\vec{\tau}_{\rm SOT} = - \dfrac{ \mu_{B}\theta_{SH} J}{e M_s h} \vec{m}\times (\vec{m}\times \vec{m}_p)
   %&\vec{\tau}_{\rm STT}   =-(\vec{\rm v}_s\cdot \vec{\nabla})\vec{\Omega}+\beta \vec{\Omega}\times (\vec{\rm v}_s\cdot \vec{\nabla})\vec{\Omega} 
\end{align}

\noindent
where $\mu_B$ and $|e|$ are the Bohr magneton and the electron charge, respectively. The parameter $\theta_{SH}$ corresponds to the spin-Hall angle, which is $\theta_{SH}=0.19$ for the Co/Pt interface \cite{Zhang2015}.  Additionally, $J$ represents the magnitude of the in-plane  current density and $\vec{m}_p$ is the normalized spin polarization vector. For the dynamic calculations, we use $\alpha = 0.1$.\\

\begin{figure}[ht!]
    \centering
    \includegraphics[width=0.9\linewidth]{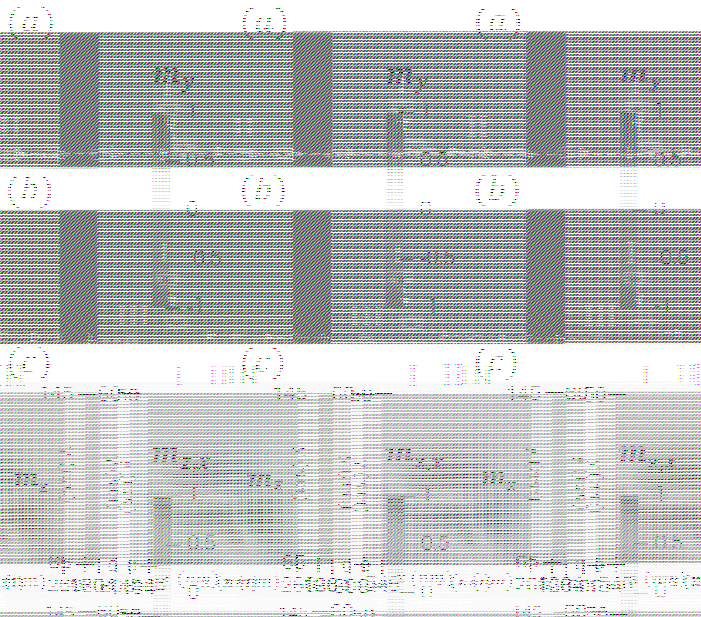}
    \caption{
    Panels (a) and (b) depict the trajectories of a magnetic bimeron under current densities of $J= -1\times 10^{11}$ A/m$^2$ and $J=- 4\times 10^{11}$ \text{A/m}$^2$, respectively. The dashed lines show the paths followed. These figures show only a reduced area of the strip, measuring $1400 \times 256$ $\text{nm}^2$ for better visualization. For further clarity on the dynamics of the bimeron, refer to Supplementary Video 1 for panel (a) and Supplementary Video 2 for panel (b).  (c) Magnetic profiles of the bimeron along the trajectories. It is observed that the bimeron decreases in size as it approaches the edge of the strip. Depending on the current value, the bimeron may be annihilated at the edge or continue along a straight trajectory at the edge as a boundary state.}
    \label{fig2}
\end{figure}

\begin{figure}[ht]
    \centering \includegraphics[width=0.9\linewidth]{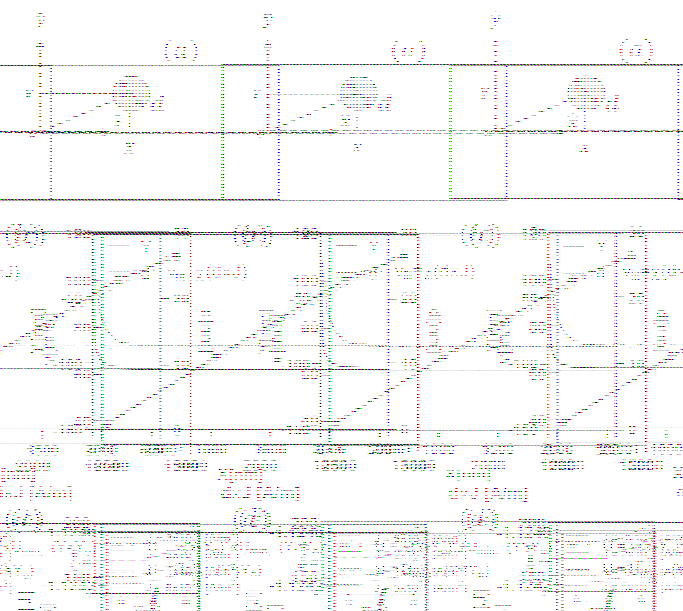}
    \caption{(a) Diagram shows the origin of the coordinate system at the initial position of the bimeron. (b) Terminal velocity of the bimeron as function of $d \times J$. The dashed line shows the fit curve with $\chi = 0.038676 [m^2/A \cdot s]$. (c) Bimeron core diameter and bimeron core position along the nano-stripe. The bimeron core diameter was calculated considering the diameter of the isocurve defined by $m_y =0$. (d)  Velocity of the bimeron for $J = -5 \times 10^{10} A/m^2$ (purple), $J =- 1 \times 10^{11} A/m^2$ (red), $J = -2 \times 10^{11} A/m^2$ (green), and $J =- 3 \times 10^{11} A/m^2$ (blue).  (e) Magnetic energy as a function of the trajectory along the width track, for $J =- 2 \times 10^{11} A/m^2$. The dashed red line denotes the proposed fitted curve with $\nu \equiv  \frac{\kappa}{2} = 7.1215  \times 10^{-5} J/m^2$.  }
    \label{fig3}
\end{figure}

\noindent{\textbf{\textit{Motion of bimeron driven by SOT -}}} Figure \ref{fig2} shows the main result of the article. For a current density below a threshold value, the bimeron moves at the edge of the strip after undergoing the hall effect, \textit{i.e.}, it does not annihilate at the strip border, as shown in Fig. \ref{fig2}.a. However, for $J$ above a critical value, the bimeron reaches the edge, being annihilated, as depicted in Fig. \ref{fig2}.b. The prevention of annihilation of the bimeron is due to its topological protection and the easy axis of uniaxial anisotropy along the y-axis (perpendicular to the axis in which the bimeron displaces). This fact can be seen in Fig. \ref{fig2}.c. In Figure \ref{fig2}.c.I., we can see the bimeron in the middle of the strip, whose magnetization along the $z$-axis  is not circularly symmetric. Two regions, characterizing the meron and antimeron cores highlighted in red and blue, can be observed: a top region with $z$-magnetization pointing upward and a bottom region with $z$-magnetization pointing downward. When bimeron annihilation does not occur ($J= -1\times 10^{11}$ A/m$^2$), it reduces its diameter $d$ until reaches the edge, see figures \ref{fig2}.a and figure \ref{fig3}.c. Upon reaching the edge, the bimeron is not annihilated because the magnetization of the strip is opposite to that of its bottom region. In this context, the top region never comes into contact with the edge of the strip, allowing the bimeron to remain in the system (see Fig. \ref{fig2}.c.II). %as a stable topological magnetic texture.
 An extra energy would be needed to rotate %part of the magnetization of 
 the bimeron and annihilate it (see Fig. \ref{fig2}.c.III). Indeed, when the current density increases in modulus ($J= -4\times 10^{11}$ A/m$^2$), the bimeron rotates so that its upper part %of the bimeron
 achieves the strip edge. The bimeron is then annihilated, changing the topology of the remaining magnetic texture, when its skyrmion number $Q\to0$, %so it is no longer a topologically protected magnetic texture, being annihilated, 
 as shown in Fig. \ref{fig2}.c.III. The phenomenology described above can be corroborated if the anisotropy easy axis rotates at an angle $\theta$, and we keep $J= -1\times 10^{11}$ A/m$^2$. Considering the easy axis as $(\sin\theta, \cos\theta, 0)$, the profile of the nucleated bimeron rotates. For $\theta \leq 30^\circ$, the bimeron still propagates through all the nanotrack. However, for $\theta \geq 45^\circ$, one observes bimeron annihilation at the strip border, which occurs because both meron-antimeron pair cores can touch the strip edge %, losing its topological protection 
 (for more details, see the Supplementary Material).

The results of the simulations reveal that the velocity of the bimeron on the strip remains constant ($v_0$) when it displaces in bulk, increasing to $v_f$  as it approaches the edge, reaching an aspect ratio $v_f/v_0=6.12$ if $J = 1\times 10^{11}$ A/m$^{2}$ and $v_f/v_0=4.59$ if $J = 2\times 10^{11}$ A/m$^2$, as shown in Figure \ref{fig3}.d. Therefore, the bimeron's velocity at the strip edge increases by over two times compared to its velocity in the bulk. Notably, we observe that this velocity increases when the bimeron moves at the edge depends linearly on the product of the bimeron diameter and the current density, see Figure \ref{fig3}.(b).  To explain this behavior, we adopt a model to capture the general properties of the bimerons by considering them as rigid structures so that their propagation can be described by Thiele's equation \cite{Thiele}, which determines the position of magnetic textures as a function of external stimuli \cite{Butner-Nat}. Considering that $\vec{R} = (X, Y)$ represents the position of the bimeron along the strip, it is possible to obtain $\vec{R}(t) = (X(t), Y(t))$ from solving the Thiele equation, given by

\begin{align}
    \vec{G}\times \vec{V} - \alpha \hat{\mathcal{D}} \vec{V} - \tau_{DL}  \vec{T}_{DL}  = \vec{\nabla} U
\end{align}

\noindent
Here, $\vec{G}=(0,0,G_z)$, is the gyrocoupling vector ($G_z= 4\pi h\,M_s  Q /\gamma$), $\hat{\mathcal{D}}$ is the dissipation dyadic matrix, $U$ is the effective potential, $h$ is the thickness of the strip, $Q$ is the skyrmion number,  $ \tau_{DL} = \dfrac{ \mu_{B}}{e M_s h}\theta_{SH}$  and $\vec{V}=d\vec{R}/dt$. The potential $U$ describes the bimeron-edge interaction, which vanishes when considering a strip with infinite width. In a strip with $w\ll h$, $U=\frac{1}{2}\kappa Y^2$ is a good approximation according to the figure \ref{fig3}.e. In this case, $\kappa$ is interpreted as a constant depending on the material parameters and the relative orientation of the bimeron with respect to the strip axis. The components $V_x(t)$ and $V_y(t)$ of the bimeron velocity are given by (see details in the Supplementary Materials).

When it achieves the positions $|Y|>|Y_c|$, the bimeron velocity along $y$-direction is
    
%    \begin{eqnarray}
%        Y(t)=-\left(Y_c-Y_\infty\right)e^{-\lambda(t-t_c)}-Y_\infty
%    \end{eqnarray}
    \begin{eqnarray}
        V_y(t) &=& \frac{dY}{dt}=\lambda \left(Y_c-Y_\infty\right)e^{-\lambda(t-t_c)}\,,
    \end{eqnarray}

    \noindent where $\lambda=\alpha\,\kappa\,\eta\, \mathcal{D}[G^2+\alpha^2\eta\, \mathcal{D}^2)]^{-1}$ and $Y_\infty$ is a current depend position of the bimeron when $t\rightarrow\infty$. $Y_c$ is the position from which the bimeron starts interacting with the nanotrack border. The bimeron reaches this position after a time $t_c$ after it starts its motion from the nanotrack center (see details in Supplemental Materials). One notices that $V_y(t\rightarrow\infty)\rightarrow0$ with a decay factor depending on $\lambda(\kappa)$, and the bimeron starts to move in a linear trajectory parallel to the strip axis with a velocity   $V_x(t\rightarrow\infty) \approx \varepsilon_x\tau_{DL}\,JR\,[{\alpha \,\eta\, \mathcal{D}^{-1}}]$, where $\varepsilon_x$ is a parameter associated with the SOT effective force along $x$ direction, appearing due to asymmetries in the solitonic profile and $\eta$ is associated with variations in the dissipative dyadic due to the bimeron deformation. It is worth noticing that $V_x\propto RJ$, and consequently, in the edge state, we obtain $V=\sqrt{V_x^2+V_y^2}\propto RJ$, which also increases with the bimeron radius and current density. Our theoretical model allows us to estimate the relation $\mathcal{V}$ between the bimeron terminal velocity and its velocity when it moves in the linear regime $V_c$ (See supplemental information). In this case, we obtain $\mathcal{V}_r=V(t\rightarrow\infty)/V_c\approx G\,R_e\,\epsilon_x\,[\mathcal{D} \alpha R\sqrt{\epsilon_x^2+\epsilon_y^2}]^{-1}$, where $R_e<R$ is the bimeron radius when it is near the strip edge. 
    In our micromagnetic simulations, we observe that $\mathcal{V}\sim G\,R_e[\mathcal{D} \alpha\sqrt{2}\,R]^{-1}\gtrsim 2$.

It is worth noticing that the results showing that the bimeron can propagate without annihilating at the edge of the strip can also be observed in a bimeron with anisotropy perpendicular to the strip surface but with an externally applied magnetic field of -120 mT in the $y$-direction (see supplementary material). In this case, the magnetic field would play the role of anisotropy in orienting the bimeron. This finding allows us to implement our results in bimeronic devices with perpendicular magnetic anisotropy.

\begin{figure}[ht!]
    \centering
    \includegraphics[width=0.85\linewidth]{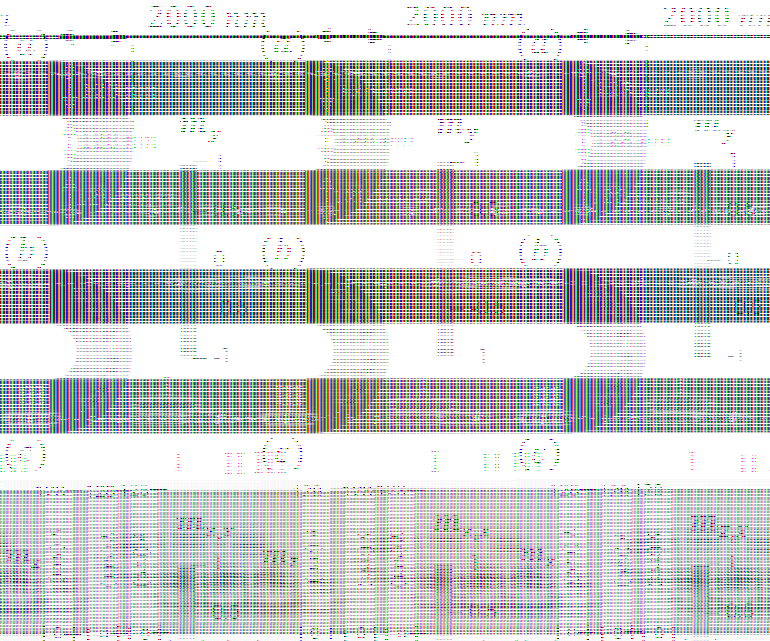}
    \caption{Isolated bimeron and bimeron chain motion in a U-shaped nanotrack.  (a) and (b) depict the trajectories of a magnetic bimeron and bimeron-chain, respectively.  The current density distribution was calculated using Comsol \cite{comsol2023}. For further clarity on the dynamics of the bimeron, refer to Supplementary Video 3 for panel (a) and Supplementary Video 4 for panel (b). (c) Magnetic profiles of the bimeron-chain along the U-shaped strip.}
    \label{fig4}
\end{figure}

\noindent{\textbf{\textit{Bimeron in magnetic memory devices - }}}For bimerons to be successfully integrated as dynamic information carriers in magnetic devices, they must demonstrate the capability to navigate through the complex, curved regions of the strip. To investigate this critical functionality, we expanded our study to include curved strips, as depicted in Fig. \ref{fig4}.a. This illustration presents a U-shaped magnetic structure characterized by a uniaxial easy axis anisotropy aligned perpendicular to the central axis of the strip. Our findings confirm that bimerons can traverse the entire system seamlessly and do so without any loss of integrity. Given the prior reports of bimeron chains \cite{Mukai2024}, our research further explored the dynamics of bimeron chain propagation along the U-shaped magnetic strip. The data robustly demonstrates that the bimeron chain maintains its stability and effectively continues its journey along the strip, undeterred by the challenging U-shaped configuration, as evidenced in Figure \ref{fig4}.b. Thus, bimeron chains are validated as viable information elements. These pivotal discoveries lay the groundwork for bimeronic racetrack memory and magnetic devices, positioning bimerons as key information carriers.

\noindent{\textbf{\textit{Final Remarks -}} } Our report addressed the edge states and stability of trajectories of isolated and chained bimerons in thin ferromagnetic strips under in-plane spin-polarized electric currents. We showed that the bimerons can propagate along the strip without annihilating due to the Hall effect when the direction of the strip easy-axis anisotropy and the electric current are perpendicular to each other. In this case, for an electric current above a threshold value, the bimeron propagates along the edge in a rectilinear trajectory along the current direction propagation. Moreover, the observed stability and dynamic properties of bimeron chains further underscore their potential as reliable information carriers. The implications of this research offer promising avenues for technological innovation in magnetic data devices.

%In conclusion, our investigation has provided significant insight into the manipulable nature of bimerons by applying uniaxial anisotropy and external magnetic fields. This manipulation has transformed bimerons into edge states on magnetic tapes, exhibiting robustness against annihilation, even in the presence of tape curvature. Moreover, the observed stability and dynamic properties of bimeron chains further underscore their potential as reliable information carriers. These findings pave the way for the development of advanced magnetic devices that utilize bimerons as fundamental information storage and transmission elements. The implications of this research are profound and offer promising avenues for technological innovation in magnetic data devices.

\noindent{\it Acknowledgements.-}  Funding is acknowledged from Fondecyt Regular 1230515 and 1220215, DICYT regular 042431AP, and Financiamiento Basal para  Centros  Cient\'ificos  y  Tecnol\'ogicos  de  Excelencia AFB220001. 
%G.S. thanks the financial support provided by ANID Subdirección de Capital Humano/Doctorado, Chile Nacional/2022-21222167. 
M.A.C. acknowledges Proyecto ANID Fondecyt Postdoctorado 3240112.  D. Gálvez-Poblete acknowledges ANID-Subdirección de Capital Humano/Doctorado Nacional/2023-21230818. V.L.C.-S. thanks to the Brazilian agencies CNPq (Grant No. 305256/2022) and Fapemig (Grant No. APQ-00648-22).

%\bibliography{apssamp}% Produces the bibliography via BibTeX.
%\bibliographystyle{apsrev}
\bibliography{References}

%apsrev4-2.bst 2019-01-14 (MD) hand-edited version of apsrev4-1.bst
%Control: key (0)
%Control: author (8) initials jnrlst
%Control: editor formatted (1) identically to author
%Control: production of article title (0) allowed
%Control: page (0) single
%Control: year (1) truncated
%Control: production of eprint (0) enabled
\begin{thebibliography}{70}%
\makeatletter
\providecommand \@ifxundefined [1]{%
 \@ifx{#1\undefined}
}%
\providecommand \@ifnum [1]{%
 \ifnum #1\expandafter \@firstoftwo
 \else \expandafter \@secondoftwo
 \fi
}%
\providecommand \@ifx [1]{%
 \ifx #1\expandafter \@firstoftwo
 \else \expandafter \@secondoftwo
 \fi
}%
\providecommand \natexlab [1]{#1}%
\providecommand \enquote  [1]{``#1''}%
\providecommand \bibnamefont  [1]{#1}%
\providecommand \bibfnamefont [1]{#1}%
\providecommand \citenamefont [1]{#1}%
\providecommand \href@noop [0]{\@secondoftwo}%
\providecommand \href [0]{\begingroup \@sanitize@url \@href}%
\providecommand \@href[1]{\@@startlink{#1}\@@href}%
\providecommand \@@href[1]{\endgroup#1\@@endlink}%
\providecommand \@sanitize@url [0]{\catcode `\\12\catcode `\$12\catcode `\&12\catcode `\#12\catcode `\^12\catcode `\_12\catcode `\%12\relax}%
\providecommand \@@startlink[1]{}%
\providecommand \@@endlink[0]{}%
\providecommand \url  [0]{\begingroup\@sanitize@url \@url }%
\providecommand \@url [1]{\endgroup\@href {#1}{\urlprefix }}%
\providecommand \urlprefix  [0]{URL }%
\providecommand \Eprint [0]{\href }%
\providecommand \doibase [0]{https://doi.org/}%
\providecommand \selectlanguage [0]{\@gobble}%
\providecommand \bibinfo  [0]{\@secondoftwo}%
\providecommand \bibfield  [0]{\@secondoftwo}%
\providecommand \translation [1]{[#1]}%
\providecommand \BibitemOpen [0]{}%
\providecommand \bibitemStop [0]{}%
\providecommand \bibitemNoStop [0]{.\EOS\space}%
\providecommand \EOS [0]{\spacefactor3000\relax}%
\providecommand \BibitemShut  [1]{\csname bibitem#1\endcsname}%
\let\auto@bib@innerbib\@empty
%</preamble>
\bibitem [{\citenamefont {Yang}\ \emph {et~al.}(2021)\citenamefont {Yang}, \citenamefont {Naaman}, \citenamefont {Paltiel},\ and\ \citenamefont {Parkin}}]{Yang2021}%
  \BibitemOpen
  \bibfield  {author} {\bibinfo {author} {\bibfnamefont {S.-H.}\ \bibnamefont {Yang}}, \bibinfo {author} {\bibfnamefont {R.}~\bibnamefont {Naaman}}, \bibinfo {author} {\bibfnamefont {Y.}~\bibnamefont {Paltiel}},\ and\ \bibinfo {author} {\bibfnamefont {S.~S.~P.}\ \bibnamefont {Parkin}},\ }\bibfield  {title} {\bibinfo {title} {Chiral spintronics},\ }\href {https://doi.org/10.1038/s42254-021-00302-9} {\bibfield  {journal} {\bibinfo  {journal} {Nature Reviews Physics}\ }\textbf {\bibinfo {volume} {3}},\ \bibinfo {pages} {328–343} (\bibinfo {year} {2021})}\BibitemShut {NoStop}%
\bibitem [{\citenamefont {Lavrijsen}\ \emph {et~al.}(2013)\citenamefont {Lavrijsen}, \citenamefont {Lee}, \citenamefont {Fern\'andez-Pacheco}, \citenamefont {Petit}, \citenamefont {Mansell},\ and\ \citenamefont {Cowburn}}]{Ref1}%
  \BibitemOpen
  \bibfield  {author} {\bibinfo {author} {\bibfnamefont {R.}~\bibnamefont {Lavrijsen}}, \bibinfo {author} {\bibfnamefont {J.-H.}\ \bibnamefont {Lee}}, \bibinfo {author} {\bibfnamefont {A.}~\bibnamefont {Fern\'andez-Pacheco}}, \bibinfo {author} {\bibfnamefont {D.~C. M.~C.}\ \bibnamefont {Petit}}, \bibinfo {author} {\bibfnamefont {R.}~\bibnamefont {Mansell}},\ and\ \bibinfo {author} {\bibfnamefont {R.~P.}\ \bibnamefont {Cowburn}},\ }\bibfield  {title} {\bibinfo {title} {Magnetic ratchet for three-dimensional spintronic memory and logic},\ }\href@noop {} {\bibfield  {journal} {\bibinfo  {journal} {Nature}\ }\textbf {\bibinfo {volume} {493}},\ \bibinfo {pages} {647} (\bibinfo {year} {2013})}\BibitemShut {NoStop}%
\bibitem [{\citenamefont {Hrkac}\ \emph {et~al.}(2011)\citenamefont {Hrkac}, \citenamefont {Dean},\ and\ \citenamefont {Allwood}}]{Ref2}%
  \BibitemOpen
  \bibfield  {author} {\bibinfo {author} {\bibfnamefont {G.}~\bibnamefont {Hrkac}}, \bibinfo {author} {\bibfnamefont {J.}~\bibnamefont {Dean}},\ and\ \bibinfo {author} {\bibfnamefont {D.}~\bibnamefont {Allwood}},\ }\bibfield  {title} {\bibinfo {title} {Nanowire spintronics for storage class memories and logic},\ }\href@noop {} {\bibfield  {journal} {\bibinfo  {journal} {Philos. Trans. R. Soc. A}\ }\textbf {\bibinfo {volume} {369}},\ \bibinfo {pages} {3214} (\bibinfo {year} {2011})}\BibitemShut {NoStop}%
\bibitem [{\citenamefont {Hirohata}\ \emph {et~al.}(2020)\citenamefont {Hirohata}, \citenamefont {Yamada}, \citenamefont {Nakatani}, \citenamefont {Prejbeanu}, \citenamefont {Di\'eny}, \citenamefont {Pirro},\ and\ \citenamefont {Hillebrands}}]{Ref3}%
  \BibitemOpen
  \bibfield  {author} {\bibinfo {author} {\bibfnamefont {A.}~\bibnamefont {Hirohata}}, \bibinfo {author} {\bibfnamefont {K.}~\bibnamefont {Yamada}}, \bibinfo {author} {\bibfnamefont {Y.}~\bibnamefont {Nakatani}}, \bibinfo {author} {\bibfnamefont {I.-L.}\ \bibnamefont {Prejbeanu}}, \bibinfo {author} {\bibfnamefont {B.}~\bibnamefont {Di\'eny}}, \bibinfo {author} {\bibfnamefont {P.}~\bibnamefont {Pirro}},\ and\ \bibinfo {author} {\bibfnamefont {B.}~\bibnamefont {Hillebrands}},\ }\bibfield  {title} {\bibinfo {title} {Review on spintronics: Principles and device applications},\ }\href@noop {} {\bibfield  {journal} {\bibinfo  {journal} {J. Magn. Mag. Mat.}\ }\textbf {\bibinfo {volume} {509}},\ \bibinfo {pages} {166711} (\bibinfo {year} {2020})}\BibitemShut {NoStop}%
\bibitem [{\citenamefont {He}\ \emph {et~al.}(2021)\citenamefont {He}, \citenamefont {Hughes}, \citenamefont {Armitage}, \citenamefont {Tokura},\ and\ \citenamefont {Wang}}]{He2021}%
  \BibitemOpen
  \bibfield  {author} {\bibinfo {author} {\bibfnamefont {Q.~L.}\ \bibnamefont {He}}, \bibinfo {author} {\bibfnamefont {T.~L.}\ \bibnamefont {Hughes}}, \bibinfo {author} {\bibfnamefont {N.~P.}\ \bibnamefont {Armitage}}, \bibinfo {author} {\bibfnamefont {Y.}~\bibnamefont {Tokura}},\ and\ \bibinfo {author} {\bibfnamefont {K.~L.}\ \bibnamefont {Wang}},\ }\bibfield  {title} {\bibinfo {title} {Topological spintronics and magnetoelectronics},\ }\href {https://doi.org/10.1038/s41563-021-01138-5} {\bibfield  {journal} {\bibinfo  {journal} {Nature Materials}\ }\textbf {\bibinfo {volume} {21}},\ \bibinfo {pages} {15–23} (\bibinfo {year} {2021})}\BibitemShut {NoStop}%
\bibitem [{\citenamefont {Parkin}\ \emph {et~al.}(2008)\citenamefont {Parkin}, \citenamefont {Hayashi},\ and\ \citenamefont {Thomas}}]{Parkin-Sci}%
  \BibitemOpen
  \bibfield  {author} {\bibinfo {author} {\bibfnamefont {S.~S.~P.}\ \bibnamefont {Parkin}}, \bibinfo {author} {\bibfnamefont {M.}~\bibnamefont {Hayashi}},\ and\ \bibinfo {author} {\bibfnamefont {L.}~\bibnamefont {Thomas}},\ }\bibfield  {title} {\bibinfo {title} {Magnetic domain-wall racetrack memory},\ }\href@noop {} {\bibfield  {journal} {\bibinfo  {journal} {Science}\ }\textbf {\bibinfo {volume} {320}},\ \bibinfo {pages} {190} (\bibinfo {year} {2008})}\BibitemShut {NoStop}%
\bibitem [{\citenamefont {Parkin}\ and\ \citenamefont {Yang}(2015)}]{Parkin-Nat}%
  \BibitemOpen
  \bibfield  {author} {\bibinfo {author} {\bibfnamefont {S.}~\bibnamefont {Parkin}}\ and\ \bibinfo {author} {\bibfnamefont {S.-H.}\ \bibnamefont {Yang}},\ }\bibfield  {title} {\bibinfo {title} {Memory on the racetrack},\ }\href@noop {} {\bibfield  {journal} {\bibinfo  {journal} {Nat. Nano.}\ }\textbf {\bibinfo {volume} {10}},\ \bibinfo {pages} {195} (\bibinfo {year} {2015})}\BibitemShut {NoStop}%
\bibitem [{\citenamefont {Sampaio}\ \emph {et~al.}(2013)\citenamefont {Sampaio}, \citenamefont {Cros}, \citenamefont {Rohart}, \citenamefont {Thiaville},\ and\ \citenamefont {Fert}}]{JSampaio2013}%
  \BibitemOpen
  \bibfield  {author} {\bibinfo {author} {\bibfnamefont {J.}~\bibnamefont {Sampaio}}, \bibinfo {author} {\bibfnamefont {V.}~\bibnamefont {Cros}}, \bibinfo {author} {\bibfnamefont {S.}~\bibnamefont {Rohart}}, \bibinfo {author} {\bibfnamefont {A.}~\bibnamefont {Thiaville}},\ and\ \bibinfo {author} {\bibfnamefont {A.}~\bibnamefont {Fert}},\ }\bibfield  {title} {\bibinfo {title} {Nucleation, stability and current-induced motion of isolated magnetic skyrmions in nanostructures},\ }\href {https://doi.org/https://doi.org/10.1038/nnano.2013.210} {\bibfield  {journal} {\bibinfo  {journal} {Nature Nanotechnol.}\ }\textbf {\bibinfo {volume} {8}},\ \bibinfo {pages} {839} (\bibinfo {year} {2013})}\BibitemShut {NoStop}%
\bibitem [{\citenamefont {Blasing}\ \emph {et~al.}(2020)\citenamefont {Blasing}, \citenamefont {Khan}, \citenamefont {Filippou}, \citenamefont {Garg}, \citenamefont {Hameed}, \citenamefont {Castrillon},\ and\ \citenamefont {Parkin}}]{Blasing2020}%
  \BibitemOpen
  \bibfield  {author} {\bibinfo {author} {\bibfnamefont {R.}~\bibnamefont {Blasing}}, \bibinfo {author} {\bibfnamefont {A.~A.}\ \bibnamefont {Khan}}, \bibinfo {author} {\bibfnamefont {P.~C.}\ \bibnamefont {Filippou}}, \bibinfo {author} {\bibfnamefont {C.}~\bibnamefont {Garg}}, \bibinfo {author} {\bibfnamefont {F.}~\bibnamefont {Hameed}}, \bibinfo {author} {\bibfnamefont {J.}~\bibnamefont {Castrillon}},\ and\ \bibinfo {author} {\bibfnamefont {S.~S.~P.}\ \bibnamefont {Parkin}},\ }\bibfield  {title} {\bibinfo {title} {Magnetic racetrack memory: From physics to the cusp of applications within a decade},\ }\href {https://doi.org/10.1109/jproc.2020.2975719} {\bibfield  {journal} {\bibinfo  {journal} {Proceedings of the IEEE}\ }\textbf {\bibinfo {volume} {108}},\ \bibinfo {pages} {1303–1321} (\bibinfo {year} {2020})}\BibitemShut {NoStop}%
\bibitem [{\citenamefont {Parkin}(2022)}]{Parkin2022}%
  \BibitemOpen
  \bibfield  {author} {\bibinfo {author} {\bibfnamefont {S.}~\bibnamefont {Parkin}},\ }\bibfield  {title} {\bibinfo {title} {Racetrack memory: a high capacity, high performance, non-volatile spintronic memory},\ }in\ \href {https://doi.org/10.1109/imw52921.2022.9779286} {\emph {\bibinfo {booktitle} {2022 IEEE International Memory Workshop (IMW)}}}\ (\bibinfo  {publisher} {IEEE},\ \bibinfo {year} {2022})\BibitemShut {NoStop}%
\bibitem [{\citenamefont {Gu}\ \emph {et~al.}(2022)\citenamefont {Gu}, \citenamefont {Guan}, \citenamefont {Hazra}, \citenamefont {Deniz}, \citenamefont {Migliorini}, \citenamefont {Zhang},\ and\ \citenamefont {Parkin}}]{Gu2022}%
  \BibitemOpen
  \bibfield  {author} {\bibinfo {author} {\bibfnamefont {K.}~\bibnamefont {Gu}}, \bibinfo {author} {\bibfnamefont {Y.}~\bibnamefont {Guan}}, \bibinfo {author} {\bibfnamefont {B.~K.}\ \bibnamefont {Hazra}}, \bibinfo {author} {\bibfnamefont {H.}~\bibnamefont {Deniz}}, \bibinfo {author} {\bibfnamefont {A.}~\bibnamefont {Migliorini}}, \bibinfo {author} {\bibfnamefont {W.}~\bibnamefont {Zhang}},\ and\ \bibinfo {author} {\bibfnamefont {S.~S.~P.}\ \bibnamefont {Parkin}},\ }\bibfield  {title} {\bibinfo {title} {Three-dimensional racetrack memory devices designed from freestanding magnetic heterostructures},\ }\href {https://doi.org/10.1038/s41565-022-01213-1} {\bibfield  {journal} {\bibinfo  {journal} {Nature Nanotechnology}\ }\textbf {\bibinfo {volume} {17}},\ \bibinfo {pages} {1065–1071} (\bibinfo {year} {2022})}\BibitemShut {NoStop}%
\bibitem [{\citenamefont {Fernández-Pacheco}\ and\ \citenamefont {Donnelly}(2022)}]{FernndezPacheco2022}%
  \BibitemOpen
  \bibfield  {author} {\bibinfo {author} {\bibfnamefont {A.}~\bibnamefont {Fernández-Pacheco}}\ and\ \bibinfo {author} {\bibfnamefont {C.}~\bibnamefont {Donnelly}},\ }\bibfield  {title} {\bibinfo {title} {The racetrack breaks free from the substrate},\ }\href {https://doi.org/10.1038/s41565-022-01206-0} {\bibfield  {journal} {\bibinfo  {journal} {Nature Nanotechnology}\ }\textbf {\bibinfo {volume} {17}},\ \bibinfo {pages} {1038–1039} (\bibinfo {year} {2022})}\BibitemShut {NoStop}%
\bibitem [{\citenamefont {Kumar}\ \emph {et~al.}(2019)\citenamefont {Kumar}, \citenamefont {Jin}, \citenamefont {Risi}, \citenamefont {Sbiaa}, \citenamefont {Lew},\ and\ \citenamefont {Piramanayagam}}]{Indian-IEEE}%
  \BibitemOpen
  \bibfield  {author} {\bibinfo {author} {\bibfnamefont {D.}~\bibnamefont {Kumar}}, \bibinfo {author} {\bibfnamefont {T.}~\bibnamefont {Jin}}, \bibinfo {author} {\bibfnamefont {S.~A.}\ \bibnamefont {Risi}}, \bibinfo {author} {\bibfnamefont {R.}~\bibnamefont {Sbiaa}}, \bibinfo {author} {\bibfnamefont {W.~S.}\ \bibnamefont {Lew}},\ and\ \bibinfo {author} {\bibfnamefont {S.~N.}\ \bibnamefont {Piramanayagam}},\ }\bibfield  {title} {\bibinfo {title} {Domain wall motion control for racetrack memory applications},\ }\href@noop {} {\bibfield  {journal} {\bibinfo  {journal} {IEEE Trans. Magn.}\ }\textbf {\bibinfo {volume} {55}},\ \bibinfo {pages} {2300708} (\bibinfo {year} {2019})}\BibitemShut {NoStop}%
\bibitem [{\citenamefont {Geng}\ and\ \citenamefont {Jin}(2017)}]{Geng-JMMM}%
  \BibitemOpen
  \bibfield  {author} {\bibinfo {author} {\bibfnamefont {L.~D.}\ \bibnamefont {Geng}}\ and\ \bibinfo {author} {\bibfnamefont {Y.~M.}\ \bibnamefont {Jin}},\ }\bibfield  {title} {\bibinfo {title} {Magnetic vortex racetrack memory},\ }\href@noop {} {\bibfield  {journal} {\bibinfo  {journal} {J. Magn. Mag. Mat.}\ }\textbf {\bibinfo {volume} {423}},\ \bibinfo {pages} {84} (\bibinfo {year} {2017})}\BibitemShut {NoStop}%
\bibitem [{\citenamefont {Tomasello}\ \emph {et~al.}(2014{\natexlab{a}})\citenamefont {Tomasello}, \citenamefont {Martinez}, \citenamefont {Zivieri}, \citenamefont {Torres}, \citenamefont {Carpentieri},\ and\ \citenamefont {Finocchio}}]{Tomasello-Sci}%
  \BibitemOpen
  \bibfield  {author} {\bibinfo {author} {\bibfnamefont {R.}~\bibnamefont {Tomasello}}, \bibinfo {author} {\bibfnamefont {E.}~\bibnamefont {Martinez}}, \bibinfo {author} {\bibfnamefont {R.}~\bibnamefont {Zivieri}}, \bibinfo {author} {\bibfnamefont {L.}~\bibnamefont {Torres}}, \bibinfo {author} {\bibfnamefont {M.}~\bibnamefont {Carpentieri}},\ and\ \bibinfo {author} {\bibfnamefont {G.}~\bibnamefont {Finocchio}},\ }\bibfield  {title} {\bibinfo {title} {A strategy for the design of skyrmion racetrack memories},\ }\href@noop {} {\bibfield  {journal} {\bibinfo  {journal} {Sci. Rep.}\ }\textbf {\bibinfo {volume} {4}},\ \bibinfo {pages} {6784} (\bibinfo {year} {2014}{\natexlab{a}})}\BibitemShut {NoStop}%
\bibitem [{\citenamefont {Tomasello}\ \emph {et~al.}(2017)\citenamefont {Tomasello}, \citenamefont {Puliafito}, \citenamefont {Martinez}, \citenamefont {Manchon}, \citenamefont {Ricci}, \citenamefont {Carpentieri},\ and\ \citenamefont {Finocchio}}]{Tomasello-JPD}%
  \BibitemOpen
  \bibfield  {author} {\bibinfo {author} {\bibfnamefont {R.}~\bibnamefont {Tomasello}}, \bibinfo {author} {\bibfnamefont {V.}~\bibnamefont {Puliafito}}, \bibinfo {author} {\bibfnamefont {E.}~\bibnamefont {Martinez}}, \bibinfo {author} {\bibfnamefont {A.}~\bibnamefont {Manchon}}, \bibinfo {author} {\bibfnamefont {M.}~\bibnamefont {Ricci}}, \bibinfo {author} {\bibfnamefont {M.}~\bibnamefont {Carpentieri}},\ and\ \bibinfo {author} {\bibfnamefont {G.}~\bibnamefont {Finocchio}},\ }\bibfield  {title} {\bibinfo {title} {Performance of synthetic antiferromagnetic racetrack memory: domain wall versus skyrmion},\ }\href@noop {} {\bibfield  {journal} {\bibinfo  {journal} {J. Phys. D: Appl. Phys.}\ }\textbf {\bibinfo {volume} {50}},\ \bibinfo {pages} {325302} (\bibinfo {year} {2017})}\BibitemShut {NoStop}%
\bibitem [{\citenamefont {Liang}\ \emph {et~al.}(2019)\citenamefont {Liang}, \citenamefont {Zhao}, \citenamefont {Shen}, \citenamefont {Xia}, \citenamefont {Zhao}, \citenamefont {Zhang},\ and\ \citenamefont {Zhou}}]{Liang-PRB}%
  \BibitemOpen
  \bibfield  {author} {\bibinfo {author} {\bibfnamefont {X.}~\bibnamefont {Liang}}, \bibinfo {author} {\bibfnamefont {G.}~\bibnamefont {Zhao}}, \bibinfo {author} {\bibfnamefont {L.}~\bibnamefont {Shen}}, \bibinfo {author} {\bibfnamefont {J.}~\bibnamefont {Xia}}, \bibinfo {author} {\bibfnamefont {L.}~\bibnamefont {Zhao}}, \bibinfo {author} {\bibfnamefont {X.}~\bibnamefont {Zhang}},\ and\ \bibinfo {author} {\bibfnamefont {Y.}~\bibnamefont {Zhou}},\ }\bibfield  {title} {\bibinfo {title} {Dynamics of an antiferromagnetic skyrmion in a racetrack with a defect},\ }\href@noop {} {\bibfield  {journal} {\bibinfo  {journal} {Phys. Rev. B}\ }\textbf {\bibinfo {volume} {100}},\ \bibinfo {pages} {144439} (\bibinfo {year} {2019})}\BibitemShut {NoStop}%
\bibitem [{\citenamefont {Zhang}\ \emph {et~al.}(2016)\citenamefont {Zhang}, \citenamefont {Xia}, \citenamefont {Zhou}, \citenamefont {Wang}, \citenamefont {Liu}, \citenamefont {Zhao},\ and\ \citenamefont {Ezawa}}]{Zhang-PRB}%
  \BibitemOpen
  \bibfield  {author} {\bibinfo {author} {\bibfnamefont {X.}~\bibnamefont {Zhang}}, \bibinfo {author} {\bibfnamefont {J.}~\bibnamefont {Xia}}, \bibinfo {author} {\bibfnamefont {Y.}~\bibnamefont {Zhou}}, \bibinfo {author} {\bibfnamefont {D.}~\bibnamefont {Wang}}, \bibinfo {author} {\bibfnamefont {X.}~\bibnamefont {Liu}}, \bibinfo {author} {\bibfnamefont {W.}~\bibnamefont {Zhao}},\ and\ \bibinfo {author} {\bibfnamefont {M.}~\bibnamefont {Ezawa}},\ }\bibfield  {title} {\bibinfo {title} {Control and manipulation of a magnetic skyrmionium in nanostructures},\ }\href@noop {} {\bibfield  {journal} {\bibinfo  {journal} {Phys. Rev. B}\ }\textbf {\bibinfo {volume} {94}},\ \bibinfo {pages} {094420} (\bibinfo {year} {2016})}\BibitemShut {NoStop}%
\bibitem [{\citenamefont {Vigo-Cotrina}\ \emph {et~al.}(2024)\citenamefont {Vigo-Cotrina}, \citenamefont {Navarro-Vilca},\ and\ \citenamefont {Urcia-Romero}}]{Cotrina-APL}%
  \BibitemOpen
  \bibfield  {author} {\bibinfo {author} {\bibfnamefont {H.}~\bibnamefont {Vigo-Cotrina}}, \bibinfo {author} {\bibfnamefont {S.}~\bibnamefont {Navarro-Vilca}},\ and\ \bibinfo {author} {\bibfnamefont {S.}~\bibnamefont {Urcia-Romero}},\ }\bibfield  {title} {\bibinfo {title} {Skyrmionium dynamics on a racetrack in the presence of a magnetic defect},\ }\href@noop {} {\bibfield  {journal} {\bibinfo  {journal} {Appl. Phys. Lett.}\ }\textbf {\bibinfo {volume} {135}},\ \bibinfo {pages} {163903} (\bibinfo {year} {2024})}\BibitemShut {NoStop}%
\bibitem [{\citenamefont {Liu}\ \emph {et~al.}(2020)\citenamefont {Liu}, \citenamefont {Hou}, \citenamefont {Han},\ and\ \citenamefont {Zang}}]{Liu-PRL}%
  \BibitemOpen
  \bibfield  {author} {\bibinfo {author} {\bibfnamefont {Y.}~\bibnamefont {Liu}}, \bibinfo {author} {\bibfnamefont {W.}~\bibnamefont {Hou}}, \bibinfo {author} {\bibfnamefont {X.}~\bibnamefont {Han}},\ and\ \bibinfo {author} {\bibfnamefont {J.}~\bibnamefont {Zang}},\ }\bibfield  {title} {\bibinfo {title} {Three-dimensional dynamics of a magnetic hopfion driven by spin transfer torque},\ }\href@noop {} {\bibfield  {journal} {\bibinfo  {journal} {Phys. Rev. Lett.}\ }\textbf {\bibinfo {volume} {124}},\ \bibinfo {pages} {127204} (\bibinfo {year} {2020})}\BibitemShut {NoStop}%
\bibitem [{\citenamefont {Wang}\ \emph {et~al.}(2019{\natexlab{a}})\citenamefont {Wang}, \citenamefont {Qaiumzadeh},\ and\ \citenamefont {Brataas}}]{Wang-PRL}%
  \BibitemOpen
  \bibfield  {author} {\bibinfo {author} {\bibfnamefont {X.~S.}\ \bibnamefont {Wang}}, \bibinfo {author} {\bibfnamefont {A.}~\bibnamefont {Qaiumzadeh}},\ and\ \bibinfo {author} {\bibfnamefont {A.}~\bibnamefont {Brataas}},\ }\bibfield  {title} {\bibinfo {title} {Current-driven dynamics of magnetic hopfions},\ }\href@noop {} {\bibfield  {journal} {\bibinfo  {journal} {Phys. Rev. Lett.}\ }\textbf {\bibinfo {volume} {123}},\ \bibinfo {pages} {147203} (\bibinfo {year} {2019}{\natexlab{a}})}\BibitemShut {NoStop}%
\bibitem [{\citenamefont {Kim}(2019)}]{Kim-PRB}%
  \BibitemOpen
  \bibfield  {author} {\bibinfo {author} {\bibfnamefont {S.~K.}\ \bibnamefont {Kim}},\ }\bibfield  {title} {\bibinfo {title} {Dynamics of bimeron skyrmions in easy-plane magnets induced by a spin supercurrent},\ }\href@noop {} {\bibfield  {journal} {\bibinfo  {journal} {Phys. Rev. B}\ }\textbf {\bibinfo {volume} {94}},\ \bibinfo {pages} {224406} (\bibinfo {year} {2019})}\BibitemShut {NoStop}%
\bibitem [{\citenamefont {Castro}\ \emph {et~al.}(2023)\citenamefont {Castro}, \citenamefont {Altbir}, \citenamefont {Galvez-Poblete}, \citenamefont {Corona}, \citenamefont {Oyarzún}, \citenamefont {Pereira}, \citenamefont {Allende},\ and\ \citenamefont {Carvalho-Santos}}]{CastroMA2023}%
  \BibitemOpen
  \bibfield  {author} {\bibinfo {author} {\bibfnamefont {M.}~\bibnamefont {Castro}}, \bibinfo {author} {\bibfnamefont {D.}~\bibnamefont {Altbir}}, \bibinfo {author} {\bibfnamefont {D.}~\bibnamefont {Galvez-Poblete}}, \bibinfo {author} {\bibfnamefont {R.}~\bibnamefont {Corona}}, \bibinfo {author} {\bibfnamefont {S.}~\bibnamefont {Oyarzún}}, \bibinfo {author} {\bibfnamefont {A.}~\bibnamefont {Pereira}}, \bibinfo {author} {\bibfnamefont {S.}~\bibnamefont {Allende}},\ and\ \bibinfo {author} {\bibfnamefont {V.}~\bibnamefont {Carvalho-Santos}},\ }\bibfield  {title} {\bibinfo {title} {Skyrmion-bimeron dynamic conversion in magnetic nanotracks},\ }\bibfield  {journal} {\bibinfo  {journal} {Phys. Rev. B}\ }\textbf {\bibinfo {volume} {108}},\ \href {https://doi.org/10.1103/PhysRevB.108.094436} {10.1103/PhysRevB.108.094436} (\bibinfo {year} {2023})\BibitemShut {NoStop}%
\bibitem [{\citenamefont {Li}\ \emph {et~al.}(2020)\citenamefont {Li}, \citenamefont {Shen}, \citenamefont {Bai}, \citenamefont {Wang}, \citenamefont {Zhang}, \citenamefont {Xia}, \citenamefont {Ezawa}, \citenamefont {Tretiakov}, \citenamefont {Xu}, \citenamefont {Mruczkiewicz}, \citenamefont {Krawczyk}, \citenamefont {Xu}, \citenamefont {Evans}, \citenamefont {Chantrell},\ and\ \citenamefont {Zhou}}]{Li-NPJ}%
  \BibitemOpen
  \bibfield  {author} {\bibinfo {author} {\bibfnamefont {X.}~\bibnamefont {Li}}, \bibinfo {author} {\bibfnamefont {L.}~\bibnamefont {Shen}}, \bibinfo {author} {\bibfnamefont {Y.}~\bibnamefont {Bai}}, \bibinfo {author} {\bibfnamefont {J.}~\bibnamefont {Wang}}, \bibinfo {author} {\bibfnamefont {X.}~\bibnamefont {Zhang}}, \bibinfo {author} {\bibfnamefont {J.}~\bibnamefont {Xia}}, \bibinfo {author} {\bibfnamefont {M.}~\bibnamefont {Ezawa}}, \bibinfo {author} {\bibfnamefont {O.~A.}\ \bibnamefont {Tretiakov}}, \bibinfo {author} {\bibfnamefont {X.}~\bibnamefont {Xu}}, \bibinfo {author} {\bibfnamefont {M.}~\bibnamefont {Mruczkiewicz}}, \bibinfo {author} {\bibfnamefont {M.}~\bibnamefont {Krawczyk}}, \bibinfo {author} {\bibfnamefont {Y.}~\bibnamefont {Xu}}, \bibinfo {author} {\bibfnamefont {R.~F.~L.}\ \bibnamefont {Evans}}, \bibinfo {author} {\bibfnamefont {R.~W.}\ \bibnamefont {Chantrell}},\ and\ \bibinfo {author} {\bibfnamefont {Y.}~\bibnamefont {Zhou}},\ }\bibfield  {title} {\bibinfo {title} {Bimeron clusters in
  chiral antiferromagnets},\ }\href@noop {} {\bibfield  {journal} {\bibinfo  {journal} {NPJ Comput. Mat.}\ }\textbf {\bibinfo {volume} {4}},\ \bibinfo {pages} {169} (\bibinfo {year} {2020})}\BibitemShut {NoStop}%
\bibitem [{\citenamefont {Liang}\ \emph {et~al.}(2023)\citenamefont {Liang}, \citenamefont {Lan}, \citenamefont {Zhao}, \citenamefont {Zelent}, \citenamefont {Krawczyk},\ and\ \citenamefont {Zhou}}]{Liang-PRB2}%
  \BibitemOpen
  \bibfield  {author} {\bibinfo {author} {\bibfnamefont {X.}~\bibnamefont {Liang}}, \bibinfo {author} {\bibfnamefont {J.}~\bibnamefont {Lan}}, \bibinfo {author} {\bibfnamefont {G.}~\bibnamefont {Zhao}}, \bibinfo {author} {\bibfnamefont {M.}~\bibnamefont {Zelent}}, \bibinfo {author} {\bibfnamefont {M.}~\bibnamefont {Krawczyk}},\ and\ \bibinfo {author} {\bibfnamefont {Y.}~\bibnamefont {Zhou}},\ }\bibfield  {title} {\bibinfo {title} {Bidirectional magnon-driven bimeron motion in ferromagnets},\ }\href@noop {} {\bibfield  {journal} {\bibinfo  {journal} {Phys. Rev. B}\ }\textbf {\bibinfo {volume} {108}},\ \bibinfo {pages} {184407} (\bibinfo {year} {2023})}\BibitemShut {NoStop}%
\bibitem [{\citenamefont {Schryer}\ and\ \citenamefont {Walker}(1974)}]{Walker}%
  \BibitemOpen
  \bibfield  {author} {\bibinfo {author} {\bibfnamefont {N.~L.}\ \bibnamefont {Schryer}}\ and\ \bibinfo {author} {\bibfnamefont {L.~R.}\ \bibnamefont {Walker}},\ }\bibfield  {title} {\bibinfo {title} {The motion of 180$^o$ domain walls in uniform dc magnetic fields},\ }\href@noop {} {\bibfield  {journal} {\bibinfo  {journal} {J. Appl. Phys.}\ }\textbf {\bibinfo {volume} {45}},\ \bibinfo {pages} {5406} (\bibinfo {year} {1974})}\BibitemShut {NoStop}%
\bibitem [{\citenamefont {Mougin}\ \emph {et~al.}(2007)\citenamefont {Mougin}, \citenamefont {Cormier}, \citenamefont {Adam}, \citenamefont {Metaxas},\ and\ \citenamefont {Ferr\'e}}]{Mougin}%
  \BibitemOpen
  \bibfield  {author} {\bibinfo {author} {\bibfnamefont {A.}~\bibnamefont {Mougin}}, \bibinfo {author} {\bibfnamefont {M.}~\bibnamefont {Cormier}}, \bibinfo {author} {\bibfnamefont {J.~P.}\ \bibnamefont {Adam}}, \bibinfo {author} {\bibfnamefont {P.~J.}\ \bibnamefont {Metaxas}},\ and\ \bibinfo {author} {\bibfnamefont {J.}~\bibnamefont {Ferr\'e}},\ }\bibfield  {title} {\bibinfo {title} {Domain wall mobility, stability and walker breakdown in magnetic nanowires},\ }\href@noop {} {\bibfield  {journal} {\bibinfo  {journal} {Europhys. Lett.}\ }\textbf {\bibinfo {volume} {78}},\ \bibinfo {pages} {57007} (\bibinfo {year} {2007})}\BibitemShut {NoStop}%
\bibitem [{\citenamefont {Lee}\ \emph {et~al.}(2007)\citenamefont {Lee}, \citenamefont {Lee}, \citenamefont {Choi}, \citenamefont {Guslienko},\ and\ \citenamefont {Kim}}]{Guslienko}%
  \BibitemOpen
  \bibfield  {author} {\bibinfo {author} {\bibfnamefont {J.-Y.}\ \bibnamefont {Lee}}, \bibinfo {author} {\bibfnamefont {K.-S.}\ \bibnamefont {Lee}}, \bibinfo {author} {\bibfnamefont {S.}~\bibnamefont {Choi}}, \bibinfo {author} {\bibfnamefont {K.~Y.}\ \bibnamefont {Guslienko}},\ and\ \bibinfo {author} {\bibfnamefont {S.-K.}\ \bibnamefont {Kim}},\ }\bibfield  {title} {\bibinfo {title} {Dynamic transformations of the internal structure of a moving domain wall in magnetic nanostripes},\ }\href@noop {} {\bibfield  {journal} {\bibinfo  {journal} {Phys. Rev. B}\ }\textbf {\bibinfo {volume} {76}},\ \bibinfo {pages} {184408} (\bibinfo {year} {2007})}\BibitemShut {NoStop}%
\bibitem [{\citenamefont {Thomas}\ \emph {et~al.}(2006)\citenamefont {Thomas}, \citenamefont {Hayashi}, \citenamefont {Jiang}, \citenamefont {Moriya}, \citenamefont {Rettner},\ and\ \citenamefont {Parkin}}]{Parkin-Nat3}%
  \BibitemOpen
  \bibfield  {author} {\bibinfo {author} {\bibfnamefont {L.}~\bibnamefont {Thomas}}, \bibinfo {author} {\bibfnamefont {M.}~\bibnamefont {Hayashi}}, \bibinfo {author} {\bibfnamefont {X.}~\bibnamefont {Jiang}}, \bibinfo {author} {\bibfnamefont {R.}~\bibnamefont {Moriya}}, \bibinfo {author} {\bibfnamefont {C.}~\bibnamefont {Rettner}},\ and\ \bibinfo {author} {\bibfnamefont {S.~S.~P.}\ \bibnamefont {Parkin}},\ }\bibfield  {title} {\bibinfo {title} {Oscillatory dependence of current-driven magnetic domain wall motion on current pulse length},\ }\href@noop {} {\bibfield  {journal} {\bibinfo  {journal} {Nature}\ }\textbf {\bibinfo {volume} {443}},\ \bibinfo {pages} {197} (\bibinfo {year} {2006})}\BibitemShut {NoStop}%
\bibitem [{\citenamefont {Yuan}\ and\ \citenamefont {Wang}(2014)}]{Yuan-PRB}%
  \BibitemOpen
  \bibfield  {author} {\bibinfo {author} {\bibfnamefont {H.~Y.}\ \bibnamefont {Yuan}}\ and\ \bibinfo {author} {\bibfnamefont {X.~R.}\ \bibnamefont {Wang}},\ }\bibfield  {title} {\bibinfo {title} {Domain wall pinning in notched nanowires},\ }\href@noop {} {\bibfield  {journal} {\bibinfo  {journal} {Phys. Rev. B}\ }\textbf {\bibinfo {volume} {89}},\ \bibinfo {pages} {054423} (\bibinfo {year} {2014})}\BibitemShut {NoStop}%
\bibitem [{\citenamefont {Woo}\ \emph {et~al.}(2017)\citenamefont {Woo}, \citenamefont {Delaney},\ and\ \citenamefont {Beach}}]{Beach-Nat}%
  \BibitemOpen
  \bibfield  {author} {\bibinfo {author} {\bibfnamefont {S.}~\bibnamefont {Woo}}, \bibinfo {author} {\bibfnamefont {T.}~\bibnamefont {Delaney}},\ and\ \bibinfo {author} {\bibfnamefont {G.~S.~D.}\ \bibnamefont {Beach}},\ }\bibfield  {title} {\bibinfo {title} {Magnetic domain wall depinning assisted by spin wave bursts},\ }\href@noop {} {\bibfield  {journal} {\bibinfo  {journal} {Nat. Phys.}\ }\textbf {\bibinfo {volume} {13}},\ \bibinfo {pages} {448} (\bibinfo {year} {2017})}\BibitemShut {NoStop}%
\bibitem [{\citenamefont {Yershov}\ \emph {et~al.}(2015)\citenamefont {Yershov}, \citenamefont {Kravchuk}, \citenamefont {Sheka},\ and\ \citenamefont {Gaididei}}]{Yershov-PRB}%
  \BibitemOpen
  \bibfield  {author} {\bibinfo {author} {\bibfnamefont {K.~V.}\ \bibnamefont {Yershov}}, \bibinfo {author} {\bibfnamefont {V.~P.}\ \bibnamefont {Kravchuk}}, \bibinfo {author} {\bibfnamefont {D.~D.}\ \bibnamefont {Sheka}},\ and\ \bibinfo {author} {\bibfnamefont {Y.}~\bibnamefont {Gaididei}},\ }\bibfield  {title} {\bibinfo {title} {Curvature-induced domain wall pinning},\ }\href@noop {} {\bibfield  {journal} {\bibinfo  {journal} {Phys. Rev. B}\ }\textbf {\bibinfo {volume} {92}},\ \bibinfo {pages} {104412} (\bibinfo {year} {2015})}\BibitemShut {NoStop}%
\bibitem [{\citenamefont {Bittencourt}\ \emph {et~al.}(2022)\citenamefont {Bittencourt}, \citenamefont {Castillo-Sep\'ulveda}, \citenamefont {Chubykalo-Fesenko}, \citenamefont {Moreno}, \citenamefont {Altbir},\ and\ \citenamefont {Carvalho-Santos}}]{Bittencourt-PRB}%
  \BibitemOpen
  \bibfield  {author} {\bibinfo {author} {\bibfnamefont {G.~H.~R.}\ \bibnamefont {Bittencourt}}, \bibinfo {author} {\bibfnamefont {S.}~\bibnamefont {Castillo-Sep\'ulveda}}, \bibinfo {author} {\bibfnamefont {O.}~\bibnamefont {Chubykalo-Fesenko}}, \bibinfo {author} {\bibfnamefont {R.}~\bibnamefont {Moreno}}, \bibinfo {author} {\bibfnamefont {D.}~\bibnamefont {Altbir}},\ and\ \bibinfo {author} {\bibfnamefont {V.~L.}\ \bibnamefont {Carvalho-Santos}},\ }\bibfield  {title} {\bibinfo {title} {Domain wall damped harmonic oscillations induced by curvature gradients in elliptical magnetic nanowires},\ }\href@noop {} {\bibfield  {journal} {\bibinfo  {journal} {Phys. Rev. B}\ }\textbf {\bibinfo {volume} {106}},\ \bibinfo {pages} {174424} (\bibinfo {year} {2022})}\BibitemShut {NoStop}%
\bibitem [{\citenamefont {Iwasaki}\ \emph {et~al.}(2013)\citenamefont {Iwasaki}, \citenamefont {Mochizuki},\ and\ \citenamefont {Nagaosa}}]{Nagaosa-Nat}%
  \BibitemOpen
  \bibfield  {author} {\bibinfo {author} {\bibfnamefont {J.}~\bibnamefont {Iwasaki}}, \bibinfo {author} {\bibfnamefont {M.}~\bibnamefont {Mochizuki}},\ and\ \bibinfo {author} {\bibfnamefont {N.}~\bibnamefont {Nagaosa}},\ }\bibfield  {title} {\bibinfo {title} {Universal current-velocity relation of skyrmion motion in chiral magnets},\ }\href@noop {} {\bibfield  {journal} {\bibinfo  {journal} {Nat. Commun.}\ }\textbf {\bibinfo {volume} {4}},\ \bibinfo {pages} {1463} (\bibinfo {year} {2013})}\BibitemShut {NoStop}%
\bibitem [{\citenamefont {Litzius}\ \emph {et~al.}(2017)\citenamefont {Litzius}, \citenamefont {Lemesh}, \citenamefont {Kr\"uger}, \citenamefont {Bassirian}, \citenamefont {Caretta}, \citenamefont {Richter}, \citenamefont {B\"uttner}, \citenamefont {Sato}, \citenamefont {Tretiakov}, \citenamefont {F\"orster}, \citenamefont {Reeve}, \citenamefont {Weigand}, \citenamefont {Bykova}, \citenamefont {Stoll}, \citenamefont {Sch\"utz}, \citenamefont {Beach4},\ and\ \citenamefont {Kl\"aui}}]{Lizius-Nat1}%
  \BibitemOpen
  \bibfield  {author} {\bibinfo {author} {\bibfnamefont {K.}~\bibnamefont {Litzius}}, \bibinfo {author} {\bibfnamefont {I.}~\bibnamefont {Lemesh}}, \bibinfo {author} {\bibfnamefont {B.}~\bibnamefont {Kr\"uger}}, \bibinfo {author} {\bibfnamefont {P.}~\bibnamefont {Bassirian}}, \bibinfo {author} {\bibfnamefont {L.}~\bibnamefont {Caretta}}, \bibinfo {author} {\bibfnamefont {K.}~\bibnamefont {Richter}}, \bibinfo {author} {\bibfnamefont {F.}~\bibnamefont {B\"uttner}}, \bibinfo {author} {\bibfnamefont {K.}~\bibnamefont {Sato}}, \bibinfo {author} {\bibfnamefont {O.~A.}\ \bibnamefont {Tretiakov}}, \bibinfo {author} {\bibfnamefont {J.}~\bibnamefont {F\"orster}}, \bibinfo {author} {\bibfnamefont {R.~M.}\ \bibnamefont {Reeve}}, \bibinfo {author} {\bibfnamefont {M.}~\bibnamefont {Weigand}}, \bibinfo {author} {\bibfnamefont {I.}~\bibnamefont {Bykova}}, \bibinfo {author} {\bibfnamefont {H.}~\bibnamefont {Stoll}}, \bibinfo {author} {\bibfnamefont {G.}~\bibnamefont {Sch\"utz}}, \bibinfo {author} {\bibfnamefont {G.~S.~D.}\
  \bibnamefont {Beach4}},\ and\ \bibinfo {author} {\bibfnamefont {M.}~\bibnamefont {Kl\"aui}},\ }\bibfield  {title} {\bibinfo {title} {Skyrmion hall effect revealed by direct time-resolved x-ray microscopy},\ }\href@noop {} {\bibfield  {journal} {\bibinfo  {journal} {Nat. Phys.}\ }\textbf {\bibinfo {volume} {13}},\ \bibinfo {pages} {170} (\bibinfo {year} {2017})}\BibitemShut {NoStop}%
\bibitem [{\citenamefont {Litzius}\ \emph {et~al.}(2020)\citenamefont {Litzius}, \citenamefont {Leliaert}, \citenamefont {Bassirian}, \citenamefont {Rodrigues}, \citenamefont {Kromin}, \citenamefont {Lemesh}, \citenamefont {Zazvorka}, \citenamefont {Lee}, \citenamefont {Mulkers}, \citenamefont {Kerber}, \citenamefont {Heinze}, \citenamefont {Keil}, \citenamefont {Reeve}, \citenamefont {Weigand}, \citenamefont {Waeyenberge}, \citenamefont {Sch\"utz}, \citenamefont {Everschor-Sitte}, \citenamefont {Beach},\ and\ \citenamefont {Kl\"aui}}]{Lizius-Nat2}%
  \BibitemOpen
  \bibfield  {author} {\bibinfo {author} {\bibfnamefont {K.}~\bibnamefont {Litzius}}, \bibinfo {author} {\bibfnamefont {J.}~\bibnamefont {Leliaert}}, \bibinfo {author} {\bibfnamefont {P.}~\bibnamefont {Bassirian}}, \bibinfo {author} {\bibfnamefont {D.}~\bibnamefont {Rodrigues}}, \bibinfo {author} {\bibfnamefont {S.}~\bibnamefont {Kromin}}, \bibinfo {author} {\bibfnamefont {I.}~\bibnamefont {Lemesh}}, \bibinfo {author} {\bibfnamefont {J.}~\bibnamefont {Zazvorka}}, \bibinfo {author} {\bibfnamefont {K.-J.}\ \bibnamefont {Lee}}, \bibinfo {author} {\bibfnamefont {J.}~\bibnamefont {Mulkers}}, \bibinfo {author} {\bibfnamefont {N.}~\bibnamefont {Kerber}}, \bibinfo {author} {\bibfnamefont {D.}~\bibnamefont {Heinze}}, \bibinfo {author} {\bibfnamefont {N.}~\bibnamefont {Keil}}, \bibinfo {author} {\bibfnamefont {R.~M.}\ \bibnamefont {Reeve}}, \bibinfo {author} {\bibfnamefont {M.}~\bibnamefont {Weigand}}, \bibinfo {author} {\bibfnamefont {B.~V.}\ \bibnamefont {Waeyenberge}}, \bibinfo {author} {\bibfnamefont
  {G.}~\bibnamefont {Sch\"utz}}, \bibinfo {author} {\bibfnamefont {K.}~\bibnamefont {Everschor-Sitte}}, \bibinfo {author} {\bibfnamefont {G.~S.~D.}\ \bibnamefont {Beach}},\ and\ \bibinfo {author} {\bibfnamefont {M.}~\bibnamefont {Kl\"aui}},\ }\bibfield  {title} {\bibinfo {title} {The role of temperature and drive current in skyrmion dynamics},\ }\href@noop {} {\bibfield  {journal} {\bibinfo  {journal} {Nat. Electron.}\ }\textbf {\bibinfo {volume} {3}},\ \bibinfo {pages} {30} (\bibinfo {year} {2020})}\BibitemShut {NoStop}%
\bibitem [{\citenamefont {Jiang}\ \emph {et~al.}(2024)\citenamefont {Jiang}, \citenamefont {Zhou}, \citenamefont {Zhang},\ and\ \citenamefont {Mochizuki}}]{Jiang-PRR}%
  \BibitemOpen
  \bibfield  {author} {\bibinfo {author} {\bibfnamefont {A.}~\bibnamefont {Jiang}}, \bibinfo {author} {\bibfnamefont {Y.}~\bibnamefont {Zhou}}, \bibinfo {author} {\bibfnamefont {X.}~\bibnamefont {Zhang}},\ and\ \bibinfo {author} {\bibfnamefont {M.}~\bibnamefont {Mochizuki}},\ }\bibfield  {title} {\bibinfo {title} {Transformation of a skyrmionium to a skyrmion through the thermal annihilation of the inner skyrmion},\ }\href@noop {} {\bibfield  {journal} {\bibinfo  {journal} {Phys. Rev. Res.}\ }\textbf {\bibinfo {volume} {6}},\ \bibinfo {pages} {013229} (\bibinfo {year} {2024})}\BibitemShut {NoStop}%
\bibitem [{\citenamefont {Yang}\ \emph {et~al.}(2023)\citenamefont {Yang}, \citenamefont {Zhao}, \citenamefont {Wu}, \citenamefont {Chu}, \citenamefont {Xu}, \citenamefont {Li}, \citenamefont {Akerman},\ and\ \citenamefont {Zhou}}]{Yang-Nat}%
  \BibitemOpen
  \bibfield  {author} {\bibinfo {author} {\bibfnamefont {S.}~\bibnamefont {Yang}}, \bibinfo {author} {\bibfnamefont {Y.}~\bibnamefont {Zhao}}, \bibinfo {author} {\bibfnamefont {K.}~\bibnamefont {Wu}}, \bibinfo {author} {\bibfnamefont {Z.}~\bibnamefont {Chu}}, \bibinfo {author} {\bibfnamefont {X.}~\bibnamefont {Xu}}, \bibinfo {author} {\bibfnamefont {X.}~\bibnamefont {Li}}, \bibinfo {author} {\bibfnamefont {J.}~\bibnamefont {Akerman}},\ and\ \bibinfo {author} {\bibfnamefont {Y.}~\bibnamefont {Zhou}},\ }\bibfield  {title} {\bibinfo {title} {Reversible conversion between skyrmions and skyrmioniums},\ }\href@noop {} {\bibfield  {journal} {\bibinfo  {journal} {Nat. Commun.}\ }\textbf {\bibinfo {volume} {14}},\ \bibinfo {pages} {3406} (\bibinfo {year} {2023})}\BibitemShut {NoStop}%
\bibitem [{\citenamefont {Sutcliffe}(2017)}]{Sutclife-PRL}%
  \BibitemOpen
  \bibfield  {author} {\bibinfo {author} {\bibfnamefont {P.}~\bibnamefont {Sutcliffe}},\ }\bibfield  {title} {\bibinfo {title} {Skyrmion knots in frustrated magnets},\ }\href@noop {} {\bibfield  {journal} {\bibinfo  {journal} {Phys. Rev. Lett}\ }\textbf {\bibinfo {volume} {118}},\ \bibinfo {pages} {247203} (\bibinfo {year} {2017})}\BibitemShut {NoStop}%
\bibitem [{\citenamefont {Sutcliffe}(2018)}]{Sutclife}%
  \BibitemOpen
  \bibfield  {author} {\bibinfo {author} {\bibfnamefont {P.}~\bibnamefont {Sutcliffe}},\ }\bibfield  {title} {\bibinfo {title} {Hopfions in chiral magnets},\ }\href@noop {} {\bibfield  {journal} {\bibinfo  {journal} {J. Phys. A: Math. Theor.}\ }\textbf {\bibinfo {volume} {51}},\ \bibinfo {pages} {375401} (\bibinfo {year} {2018})}\BibitemShut {NoStop}%
\bibitem [{\citenamefont {Wang}\ \emph {et~al.}(2019{\natexlab{b}})\citenamefont {Wang}, \citenamefont {Qaiumzadeh},\ and\ \citenamefont {Brataas}}]{Brataas}%
  \BibitemOpen
  \bibfield  {author} {\bibinfo {author} {\bibfnamefont {X.-S.}\ \bibnamefont {Wang}}, \bibinfo {author} {\bibfnamefont {A.}~\bibnamefont {Qaiumzadeh}},\ and\ \bibinfo {author} {\bibfnamefont {A.}~\bibnamefont {Brataas}},\ }\bibfield  {title} {\bibinfo {title} {Current-driven dynamics of magnetic hopfions},\ }\href@noop {} {\bibfield  {journal} {\bibinfo  {journal} {Phys. Rev. Lett}\ }\textbf {\bibinfo {volume} {123}},\ \bibinfo {pages} {147203} (\bibinfo {year} {2019}{\natexlab{b}})}\BibitemShut {NoStop}%
\bibitem [{\citenamefont {Liu}\ \emph {et~al.}(2022)\citenamefont {Liu}, \citenamefont {Watanabe},\ and\ \citenamefont {Nagaosa}}]{Nagaosa-PRL}%
  \BibitemOpen
  \bibfield  {author} {\bibinfo {author} {\bibfnamefont {Y.}~\bibnamefont {Liu}}, \bibinfo {author} {\bibfnamefont {H.}~\bibnamefont {Watanabe}},\ and\ \bibinfo {author} {\bibfnamefont {N.}~\bibnamefont {Nagaosa}},\ }\bibfield  {title} {\bibinfo {title} {Emergent magnetomultipoles and nonlinear responses of a magnetic hopfion},\ }\href@noop {} {\bibfield  {journal} {\bibinfo  {journal} {Phys. Rev. Lett}\ }\textbf {\bibinfo {volume} {129}},\ \bibinfo {pages} {267201} (\bibinfo {year} {2022})}\BibitemShut {NoStop}%
\bibitem [{\citenamefont {Toscano}\ \emph {et~al.}(2020)\citenamefont {Toscano}, \citenamefont {Mendonça}, \citenamefont {Miranda}, \citenamefont {{de Araujo}}, \citenamefont {Sato}, \citenamefont {Coura},\ and\ \citenamefont {Leonel}}]{Toscano2020}%
  \BibitemOpen
  \bibfield  {author} {\bibinfo {author} {\bibfnamefont {D.}~\bibnamefont {Toscano}}, \bibinfo {author} {\bibfnamefont {J.}~\bibnamefont {Mendonça}}, \bibinfo {author} {\bibfnamefont {A.}~\bibnamefont {Miranda}}, \bibinfo {author} {\bibfnamefont {C.}~\bibnamefont {{de Araujo}}}, \bibinfo {author} {\bibfnamefont {F.}~\bibnamefont {Sato}}, \bibinfo {author} {\bibfnamefont {P.}~\bibnamefont {Coura}},\ and\ \bibinfo {author} {\bibfnamefont {S.}~\bibnamefont {Leonel}},\ }\bibfield  {title} {\bibinfo {title} {Suppression of the skyrmion hall effect in planar nanomagnets by the magnetic properties engineering: Skyrmion transport on nanotracks with magnetic strips},\ }\href@noop {} {\bibfield  {journal} {\bibinfo  {journal} {J. Magn. Mag. Mat.}\ }\textbf {\bibinfo {volume} {504}},\ \bibinfo {pages} {166655} (\bibinfo {year} {2020})}\BibitemShut {NoStop}%
\bibitem [{\citenamefont {Zhang}\ \emph {et~al.}(2017)\citenamefont {Zhang}, \citenamefont {Luo}, \citenamefont {Yan}, \citenamefont {Ou-Yang}, \citenamefont {Yang}, \citenamefont {Chen}, \citenamefont {Zhu},\ and\ \citenamefont {You}}]{Zhang-Nanoscale}%
  \BibitemOpen
  \bibfield  {author} {\bibinfo {author} {\bibfnamefont {Y.}~\bibnamefont {Zhang}}, \bibinfo {author} {\bibfnamefont {S.}~\bibnamefont {Luo}}, \bibinfo {author} {\bibfnamefont {B.}~\bibnamefont {Yan}}, \bibinfo {author} {\bibfnamefont {J.}~\bibnamefont {Ou-Yang}}, \bibinfo {author} {\bibfnamefont {X.}~\bibnamefont {Yang}}, \bibinfo {author} {\bibfnamefont {S.}~\bibnamefont {Chen}}, \bibinfo {author} {\bibfnamefont {B.}~\bibnamefont {Zhu}},\ and\ \bibinfo {author} {\bibfnamefont {L.}~\bibnamefont {You}},\ }\bibfield  {title} {\bibinfo {title} {Magnetic skyrmions without the skyrmion hall effect in a magnetic nanotrack with perpendicular anisotropy},\ }\href@noop {} {\bibfield  {journal} {\bibinfo  {journal} {Nanoscale}\ }\textbf {\bibinfo {volume} {9}},\ \bibinfo {pages} {10212} (\bibinfo {year} {2017})}\BibitemShut {NoStop}%
\bibitem [{\citenamefont {Guo}\ \emph {et~al.}(2022)\citenamefont {Guo}, \citenamefont {Hou}, \citenamefont {Zhang}, \citenamefont {Pong},\ and\ \citenamefont {Zhou}}]{Guo2022}%
  \BibitemOpen
  \bibfield  {author} {\bibinfo {author} {\bibfnamefont {J.}~\bibnamefont {Guo}}, \bibinfo {author} {\bibfnamefont {Y.}~\bibnamefont {Hou}}, \bibinfo {author} {\bibfnamefont {X.}~\bibnamefont {Zhang}}, \bibinfo {author} {\bibfnamefont {P.~W.}\ \bibnamefont {Pong}},\ and\ \bibinfo {author} {\bibfnamefont {Y.}~\bibnamefont {Zhou}},\ }\bibfield  {title} {\bibinfo {title} {Elimination of the skyrmion hall effect by tuning perpendicular magnetic anisotropy and spin polarization angle},\ }\href@noop {} {\bibfield  {journal} {\bibinfo  {journal} {Phys. Lett. A}\ }\textbf {\bibinfo {volume} {456}},\ \bibinfo {pages} {128497} (\bibinfo {year} {2022})}\BibitemShut {NoStop}%
\bibitem [{\citenamefont {Purnama}\ \emph {et~al.}(2015)\citenamefont {Purnama}, \citenamefont {Gan}, \citenamefont {Wong},\ and\ \citenamefont {Lew}}]{Purnama}%
  \BibitemOpen
  \bibfield  {author} {\bibinfo {author} {\bibfnamefont {I.}~\bibnamefont {Purnama}}, \bibinfo {author} {\bibfnamefont {W.~L.}\ \bibnamefont {Gan}}, \bibinfo {author} {\bibfnamefont {D.~W.}\ \bibnamefont {Wong}},\ and\ \bibinfo {author} {\bibfnamefont {W.~S.}\ \bibnamefont {Lew}},\ }\bibfield  {title} {\bibinfo {title} {Guided current-induced skyrmion motion in 1d potential well},\ }\href@noop {} {\bibfield  {journal} {\bibinfo  {journal} {Sci. Rep.}\ }\textbf {\bibinfo {volume} {5}},\ \bibinfo {pages} {10620} (\bibinfo {year} {2015})}\BibitemShut {NoStop}%
\bibitem [{\citenamefont {Jin}\ \emph {et~al.}(2020)\citenamefont {Jin}, \citenamefont {Ma}, \citenamefont {Song}, \citenamefont {Xia}, \citenamefont {Wang}, \citenamefont {Zhang}, \citenamefont {Zeng}, \citenamefont {Wang},\ and\ \citenamefont {Liu}}]{Jin-NPJ}%
  \BibitemOpen
  \bibfield  {author} {\bibinfo {author} {\bibfnamefont {C.}~\bibnamefont {Jin}}, \bibinfo {author} {\bibfnamefont {Y.}~\bibnamefont {Ma}}, \bibinfo {author} {\bibfnamefont {C.}~\bibnamefont {Song}}, \bibinfo {author} {\bibfnamefont {H.}~\bibnamefont {Xia}}, \bibinfo {author} {\bibfnamefont {J.}~\bibnamefont {Wang}}, \bibinfo {author} {\bibfnamefont {C.}~\bibnamefont {Zhang}}, \bibinfo {author} {\bibfnamefont {Z.}~\bibnamefont {Zeng}}, \bibinfo {author} {\bibfnamefont {J.}~\bibnamefont {Wang}},\ and\ \bibinfo {author} {\bibfnamefont {Q.}~\bibnamefont {Liu}},\ }\bibfield  {title} {\bibinfo {title} {High-frequency spin transfer nano-oscillator based on the motion of skyrmions in an annular groove},\ }\href@noop {} {\bibfield  {journal} {\bibinfo  {journal} {New J. Phys.}\ }\textbf {\bibinfo {volume} {22}},\ \bibinfo {pages} {033001} (\bibinfo {year} {2020})}\BibitemShut {NoStop}%
\bibitem [{\citenamefont {G\"obel}\ \emph {et~al.}(2019)\citenamefont {G\"obel}, \citenamefont {Mook}, \citenamefont {Henk}, \citenamefont {Mertig},\ and\ \citenamefont {Tretiakov}}]{Gobel-PRB}%
  \BibitemOpen
  \bibfield  {author} {\bibinfo {author} {\bibfnamefont {B.}~\bibnamefont {G\"obel}}, \bibinfo {author} {\bibfnamefont {A.}~\bibnamefont {Mook}}, \bibinfo {author} {\bibfnamefont {J.}~\bibnamefont {Henk}}, \bibinfo {author} {\bibfnamefont {I.}~\bibnamefont {Mertig}},\ and\ \bibinfo {author} {\bibfnamefont {O.~A.}\ \bibnamefont {Tretiakov}},\ }\bibfield  {title} {\bibinfo {title} {Magnetic bimerons as skyrmion analogues in in-plane magnets},\ }\href@noop {} {\bibfield  {journal} {\bibinfo  {journal} {Phys. Rev. B}\ }\textbf {\bibinfo {volume} {99}},\ \bibinfo {pages} {060407(R)} (\bibinfo {year} {2019})}\BibitemShut {NoStop}%
\bibitem [{\citenamefont {Amaral}\ \emph {et~al.}(2009)\citenamefont {Amaral}, \citenamefont {Silva}, \citenamefont {Pereira},\ and\ \citenamefont {Moura-Melo}}]{Silva-PRB}%
  \BibitemOpen
  \bibfield  {author} {\bibinfo {author} {\bibfnamefont {M.~A.}\ \bibnamefont {Amaral}}, \bibinfo {author} {\bibfnamefont {R.~L.}\ \bibnamefont {Silva}}, \bibinfo {author} {\bibfnamefont {A.~R.}\ \bibnamefont {Pereira}},\ and\ \bibinfo {author} {\bibfnamefont {W.~A.}\ \bibnamefont {Moura-Melo}},\ }\bibfield  {title} {\bibinfo {title} {Discrete double core skyrmions in magnetic thin films},\ }\href@noop {} {\bibfield  {journal} {\bibinfo  {journal} {J. Magn. Mag. Mat}\ }\textbf {\bibinfo {volume} {321}},\ \bibinfo {pages} {3360} (\bibinfo {year} {2009})}\BibitemShut {NoStop}%
\bibitem [{\citenamefont {Iakovlev}\ \emph {et~al.}(2018)\citenamefont {Iakovlev}, \citenamefont {Sotnikov},\ and\ \citenamefont {Mazurenko}}]{Yakolev-PRB}%
  \BibitemOpen
  \bibfield  {author} {\bibinfo {author} {\bibfnamefont {I.~A.}\ \bibnamefont {Iakovlev}}, \bibinfo {author} {\bibfnamefont {O.~M.}\ \bibnamefont {Sotnikov}},\ and\ \bibinfo {author} {\bibfnamefont {V.~V.}\ \bibnamefont {Mazurenko}},\ }\bibfield  {title} {\bibinfo {title} {Bimeron nanoconfined design},\ }\href@noop {} {\bibfield  {journal} {\bibinfo  {journal} {Phys. Rev. B}\ }\textbf {\bibinfo {volume} {97}},\ \bibinfo {pages} {184415} (\bibinfo {year} {2018})}\BibitemShut {NoStop}%
\bibitem [{\citenamefont {Rosales}\ \emph {et~al.}(2023)\citenamefont {Rosales}, \citenamefont {Albarrac\'{\i}n}, \citenamefont {Pujol},\ and\ \citenamefont {Jaubert}}]{Rosales2023}%
  \BibitemOpen
  \bibfield  {author} {\bibinfo {author} {\bibfnamefont {H.~D.}\ \bibnamefont {Rosales}}, \bibinfo {author} {\bibfnamefont {F.~A.~G.}\ \bibnamefont {Albarrac\'{\i}n}}, \bibinfo {author} {\bibfnamefont {P.}~\bibnamefont {Pujol}},\ and\ \bibinfo {author} {\bibfnamefont {L.~D.~C.}\ \bibnamefont {Jaubert}},\ }\bibfield  {title} {\bibinfo {title} {Skyrmion fluid and bimeron glass protected by a chiral spin liquid on a kagome lattice},\ }\href {https://doi.org/10.1103/PhysRevLett.130.106703} {\bibfield  {journal} {\bibinfo  {journal} {Phys. Rev. Lett.}\ }\textbf {\bibinfo {volume} {130}},\ \bibinfo {pages} {106703} (\bibinfo {year} {2023})}\BibitemShut {NoStop}%
\bibitem [{\citenamefont {Nagase}\ \emph {et~al.}(2021{\natexlab{a}})\citenamefont {Nagase}, \citenamefont {So}, \citenamefont {Yasui}, \citenamefont {Ishida}, \citenamefont {Yoshida}, \citenamefont {Tanaka}, \citenamefont {Saitoh}, \citenamefont {Ikarashi}, \citenamefont {Kawaguchi}, \citenamefont {Kuwahara},\ and\ \citenamefont {Nagao}}]{Nagase2021}%
  \BibitemOpen
  \bibfield  {author} {\bibinfo {author} {\bibfnamefont {T.}~\bibnamefont {Nagase}}, \bibinfo {author} {\bibfnamefont {Y.-G.}\ \bibnamefont {So}}, \bibinfo {author} {\bibfnamefont {H.}~\bibnamefont {Yasui}}, \bibinfo {author} {\bibfnamefont {T.}~\bibnamefont {Ishida}}, \bibinfo {author} {\bibfnamefont {H.~K.}\ \bibnamefont {Yoshida}}, \bibinfo {author} {\bibfnamefont {Y.}~\bibnamefont {Tanaka}}, \bibinfo {author} {\bibfnamefont {K.}~\bibnamefont {Saitoh}}, \bibinfo {author} {\bibfnamefont {N.}~\bibnamefont {Ikarashi}}, \bibinfo {author} {\bibfnamefont {Y.}~\bibnamefont {Kawaguchi}}, \bibinfo {author} {\bibfnamefont {M.}~\bibnamefont {Kuwahara}},\ and\ \bibinfo {author} {\bibfnamefont {M.}~\bibnamefont {Nagao}},\ }\bibfield  {title} {\bibinfo {title} {Observation of domain wall bimerons in chiral magnets},\ }\bibfield  {journal} {\bibinfo  {journal} {Nature Communications}\ }\textbf {\bibinfo {volume} {12}},\ \href {https://doi.org/10.1038/s41467-021-23845-y} {10.1038/s41467-021-23845-y} (\bibinfo {year}
  {2021}{\natexlab{a}})\BibitemShut {NoStop}%
\bibitem [{\citenamefont {Ohara}\ \emph {et~al.}(2022)\citenamefont {Ohara}, \citenamefont {Zhang}, \citenamefont {Chen}, \citenamefont {Kato}, \citenamefont {Xia}, \citenamefont {Ezawa}, \citenamefont {Tretiakov}, \citenamefont {Hou}, \citenamefont {Zhou}, \citenamefont {Zhao}, \citenamefont {Yang},\ and\ \citenamefont {Liu}}]{Ohara2022}%
  \BibitemOpen
  \bibfield  {author} {\bibinfo {author} {\bibfnamefont {K.}~\bibnamefont {Ohara}}, \bibinfo {author} {\bibfnamefont {X.}~\bibnamefont {Zhang}}, \bibinfo {author} {\bibfnamefont {Y.}~\bibnamefont {Chen}}, \bibinfo {author} {\bibfnamefont {S.}~\bibnamefont {Kato}}, \bibinfo {author} {\bibfnamefont {J.}~\bibnamefont {Xia}}, \bibinfo {author} {\bibfnamefont {M.}~\bibnamefont {Ezawa}}, \bibinfo {author} {\bibfnamefont {O.~A.}\ \bibnamefont {Tretiakov}}, \bibinfo {author} {\bibfnamefont {Z.}~\bibnamefont {Hou}}, \bibinfo {author} {\bibfnamefont {Y.}~\bibnamefont {Zhou}}, \bibinfo {author} {\bibfnamefont {G.}~\bibnamefont {Zhao}}, \bibinfo {author} {\bibfnamefont {J.}~\bibnamefont {Yang}},\ and\ \bibinfo {author} {\bibfnamefont {X.}~\bibnamefont {Liu}},\ }\bibfield  {title} {\bibinfo {title} {Reversible transformation between isolated skyrmions and bimerons},\ }\href {https://doi.org/10.1021/acs.nanolett.2c03106} {\bibfield  {journal} {\bibinfo  {journal} {Nano Letters}\ }\textbf {\bibinfo {volume} {22}},\ \bibinfo
  {pages} {8559} (\bibinfo {year} {2022})},\ \bibinfo {note} {pMID: 36259745}\BibitemShut {NoStop}%
\bibitem [{\citenamefont {Amari}\ \emph {et~al.}(2024)\citenamefont {Amari}, \citenamefont {Ross},\ and\ \citenamefont {Nitta}}]{Amari2024}%
  \BibitemOpen
  \bibfield  {author} {\bibinfo {author} {\bibfnamefont {Y.}~\bibnamefont {Amari}}, \bibinfo {author} {\bibfnamefont {C.}~\bibnamefont {Ross}},\ and\ \bibinfo {author} {\bibfnamefont {M.}~\bibnamefont {Nitta}},\ }\bibfield  {title} {\bibinfo {title} {Domain-wall skyrmion chain and domain-wall bimerons in chiral magnets},\ }\href {https://doi.org/10.1103/PhysRevB.109.104426} {\bibfield  {journal} {\bibinfo  {journal} {Phys. Rev. B}\ }\textbf {\bibinfo {volume} {109}},\ \bibinfo {pages} {104426} (\bibinfo {year} {2024})}\BibitemShut {NoStop}%
\bibitem [{\citenamefont {Yu}\ \emph {et~al.}(2023)\citenamefont {Yu}, \citenamefont {Kanazawa}, \citenamefont {Zhang}, \citenamefont {Takahashi}, \citenamefont {Iakoubovskii}, \citenamefont {Nakajima}, \citenamefont {Tanigaki}, \citenamefont {Mochizuki},\ and\ \citenamefont {Tokura}}]{Yu2023}%
  \BibitemOpen
  \bibfield  {author} {\bibinfo {author} {\bibfnamefont {X.}~\bibnamefont {Yu}}, \bibinfo {author} {\bibfnamefont {N.}~\bibnamefont {Kanazawa}}, \bibinfo {author} {\bibfnamefont {X.}~\bibnamefont {Zhang}}, \bibinfo {author} {\bibfnamefont {Y.}~\bibnamefont {Takahashi}}, \bibinfo {author} {\bibfnamefont {K.~V.}\ \bibnamefont {Iakoubovskii}}, \bibinfo {author} {\bibfnamefont {K.}~\bibnamefont {Nakajima}}, \bibinfo {author} {\bibfnamefont {T.}~\bibnamefont {Tanigaki}}, \bibinfo {author} {\bibfnamefont {M.}~\bibnamefont {Mochizuki}},\ and\ \bibinfo {author} {\bibfnamefont {Y.}~\bibnamefont {Tokura}},\ }\bibfield  {title} {\bibinfo {title} {Spontaneous vortex‐antivortex pairs and their topological transitions in a chiral‐lattice magnet},\ }\bibfield  {journal} {\bibinfo  {journal} {Advanced Materials}\ }\textbf {\bibinfo {volume} {36}},\ \href {https://doi.org/10.1002/adma.202306441} {10.1002/adma.202306441} (\bibinfo {year} {2023})\BibitemShut {NoStop}%
\bibitem [{\citenamefont {Gao}\ \emph {et~al.}(2019)\citenamefont {Gao}, \citenamefont {Je}, \citenamefont {Choi}, \citenamefont {Yang}, \citenamefont {Wang}, \citenamefont {Lee}, \citenamefont {HAn}, \citenamefont {Lee}, \citenamefont {Chao}, \citenamefont {Hwang}, \citenamefont {Li},\ and\ \citenamefont {Qiu}}]{GaoN2019}%
  \BibitemOpen
  \bibfield  {author} {\bibinfo {author} {\bibfnamefont {N.}~\bibnamefont {Gao}}, \bibinfo {author} {\bibfnamefont {S.-G.}\ \bibnamefont {Je}}, \bibinfo {author} {\bibfnamefont {J.}~\bibnamefont {Choi}}, \bibinfo {author} {\bibfnamefont {M.}~\bibnamefont {Yang}}, \bibinfo {author} {\bibfnamefont {T.}~\bibnamefont {Wang}}, \bibinfo {author} {\bibfnamefont {S.}~\bibnamefont {Lee}}, \bibinfo {author} {\bibfnamefont {H.-K.}\ \bibnamefont {HAn}}, \bibinfo {author} {\bibfnamefont {K.-S.}\ \bibnamefont {Lee}}, \bibinfo {author} {\bibfnamefont {W.}~\bibnamefont {Chao}}, \bibinfo {author} {\bibfnamefont {C.}~\bibnamefont {Hwang}}, \bibinfo {author} {\bibfnamefont {J.}~\bibnamefont {Li}},\ and\ \bibinfo {author} {\bibfnamefont {Z.}~\bibnamefont {Qiu}},\ }\bibfield  {title} {\bibinfo {title} {Creation and anihilation of topological meron pairs in in-plane magnetized films},\ }\bibfield  {journal} {\bibinfo  {journal} {Nat. Commun.}\ }\textbf {\bibinfo {volume} {10}},\ \href
  {https://doi.org/doi.org/10.1038/s41467-019-13642-z} {doi.org/10.1038/s41467-019-13642-z} (\bibinfo {year} {2019})\BibitemShut {NoStop}%
\bibitem [{\citenamefont {Nagase}\ \emph {et~al.}(2021{\natexlab{b}})\citenamefont {Nagase}, \citenamefont {So}, \citenamefont {Yasui}, \citenamefont {Ishida}, \citenamefont {Yoshida}, \citenamefont {Tanaka}, \citenamefont {Saitoh}, \citenamefont {Ikarashi}, \citenamefont {Kawaguchi}, \citenamefont {Kuwahara},\ and\ \citenamefont {Masahiro}}]{NagaseT2021}%
  \BibitemOpen
  \bibfield  {author} {\bibinfo {author} {\bibfnamefont {T.}~\bibnamefont {Nagase}}, \bibinfo {author} {\bibfnamefont {Y.-G.}\ \bibnamefont {So}}, \bibinfo {author} {\bibfnamefont {H.}~\bibnamefont {Yasui}}, \bibinfo {author} {\bibfnamefont {T.}~\bibnamefont {Ishida}}, \bibinfo {author} {\bibfnamefont {H.~K.}\ \bibnamefont {Yoshida}}, \bibinfo {author} {\bibfnamefont {Y.}~\bibnamefont {Tanaka}}, \bibinfo {author} {\bibfnamefont {K.}~\bibnamefont {Saitoh}}, \bibinfo {author} {\bibfnamefont {N.}~\bibnamefont {Ikarashi}}, \bibinfo {author} {\bibfnamefont {Y.}~\bibnamefont {Kawaguchi}}, \bibinfo {author} {\bibfnamefont {M.}~\bibnamefont {Kuwahara}},\ and\ \bibinfo {author} {\bibfnamefont {N.}~\bibnamefont {Masahiro}},\ }\bibfield  {title} {\bibinfo {title} {Observation of domain wall bimerons in chiral magnets},\ }\bibfield  {journal} {\bibinfo  {journal} {Nat. Commun.}\ }\textbf {\bibinfo {volume} {12}},\ \href {https://doi.org/doi.org/10.1038/s41467-021-23845-y} {doi.org/10.1038/s41467-021-23845-y} (\bibinfo
  {year} {2021}{\natexlab{b}})\BibitemShut {NoStop}%
\bibitem [{\citenamefont {Zarzuela}\ \emph {et~al.}(2020)\citenamefont {Zarzuela}, \citenamefont {Bharadwaj}, \citenamefont {Kim}, \citenamefont {Sinova},\ and\ \citenamefont {Everschor-Sitte}}]{Zarzuela2020}%
  \BibitemOpen
  \bibfield  {author} {\bibinfo {author} {\bibfnamefont {R.}~\bibnamefont {Zarzuela}}, \bibinfo {author} {\bibfnamefont {V.~K.}\ \bibnamefont {Bharadwaj}}, \bibinfo {author} {\bibfnamefont {K.-W.}\ \bibnamefont {Kim}}, \bibinfo {author} {\bibfnamefont {J.}~\bibnamefont {Sinova}},\ and\ \bibinfo {author} {\bibfnamefont {K.}~\bibnamefont {Everschor-Sitte}},\ }\bibfield  {title} {\bibinfo {title} {Stability and dynamics of in-plane skyrmions in collinear ferromagnets},\ }\href@noop {} {\bibfield  {journal} {\bibinfo  {journal} {Phys. Rev. B}\ }\textbf {\bibinfo {volume} {101}},\ \bibinfo {pages} {054405} (\bibinfo {year} {2020})}\BibitemShut {NoStop}%
\bibitem [{\citenamefont {Ara\'ujo}\ \emph {et~al.}(2020)\citenamefont {Ara\'ujo}, \citenamefont {Lopes}, \citenamefont {Carvalho-Santos}, \citenamefont {Pereira}, \citenamefont {Silva}, \citenamefont {Silva},\ and\ \citenamefont {Altbir}}]{Alane}%
  \BibitemOpen
  \bibfield  {author} {\bibinfo {author} {\bibfnamefont {A.~S.}\ \bibnamefont {Ara\'ujo}}, \bibinfo {author} {\bibfnamefont {R.~J.~C.}\ \bibnamefont {Lopes}}, \bibinfo {author} {\bibfnamefont {V.~L.}\ \bibnamefont {Carvalho-Santos}}, \bibinfo {author} {\bibfnamefont {A.~R.}\ \bibnamefont {Pereira}}, \bibinfo {author} {\bibfnamefont {R.~L.}\ \bibnamefont {Silva}}, \bibinfo {author} {\bibfnamefont {R.~C.}\ \bibnamefont {Silva}},\ and\ \bibinfo {author} {\bibfnamefont {D.}~\bibnamefont {Altbir}},\ }\bibfield  {title} {\bibinfo {title} {Typical skyrmions versus bimerons: A long-distance competition in ferromagnetic racetracks},\ }\href@noop {} {\bibfield  {journal} {\bibinfo  {journal} {Phys. Rev. B}\ }\textbf {\bibinfo {volume} {102}},\ \bibinfo {pages} {104409} (\bibinfo {year} {2020})}\BibitemShut {NoStop}%
\bibitem [{\citenamefont {Shen}\ \emph {et~al.}(2020)\citenamefont {Shen}, \citenamefont {Li}, \citenamefont {Xia}, \citenamefont {Qiu}, \citenamefont {Zhang}, \citenamefont {Tretiakov}, \citenamefont {Ezawa},\ and\ \citenamefont {Zhou}}]{Shen-PRB}%
  \BibitemOpen
  \bibfield  {author} {\bibinfo {author} {\bibfnamefont {L.}~\bibnamefont {Shen}}, \bibinfo {author} {\bibfnamefont {X.}~\bibnamefont {Li}}, \bibinfo {author} {\bibfnamefont {J.}~\bibnamefont {Xia}}, \bibinfo {author} {\bibfnamefont {L.}~\bibnamefont {Qiu}}, \bibinfo {author} {\bibfnamefont {X.}~\bibnamefont {Zhang}}, \bibinfo {author} {\bibfnamefont {O.~A.}\ \bibnamefont {Tretiakov}}, \bibinfo {author} {\bibfnamefont {M.}~\bibnamefont {Ezawa}},\ and\ \bibinfo {author} {\bibfnamefont {Y.}~\bibnamefont {Zhou}},\ }\bibfield  {title} {\bibinfo {title} {Dynamics of ferromagnetic bimerons driven by spin currents and magnetic fields},\ }\href@noop {} {\bibfield  {journal} {\bibinfo  {journal} {Phys. Rev. B}\ }\textbf {\bibinfo {volume} {102}},\ \bibinfo {pages} {104427} (\bibinfo {year} {2020})}\BibitemShut {NoStop}%
\bibitem [{\citenamefont {Murooka}\ \emph {et~al.}(2020)\citenamefont {Murooka}, \citenamefont {Leonov}, \citenamefont {Inoue},\ and\ \citenamefont {i.~Ohe}}]{Leonov}%
  \BibitemOpen
  \bibfield  {author} {\bibinfo {author} {\bibfnamefont {R.}~\bibnamefont {Murooka}}, \bibinfo {author} {\bibfnamefont {A.~O.}\ \bibnamefont {Leonov}}, \bibinfo {author} {\bibfnamefont {K.}~\bibnamefont {Inoue}},\ and\ \bibinfo {author} {\bibfnamefont {J.}~\bibnamefont {i.~Ohe}},\ }\bibfield  {title} {\bibinfo {title} {Current-induced shuttlecock-like movement of non-axisymmetric chiral skyrmions},\ }\href {https://doi.org/10.1038/s41598-019-56791-3} {\bibfield  {journal} {\bibinfo  {journal} {Sci. Rep.}\ }\textbf {\bibinfo {volume} {10}} (\bibinfo {year} {2020})}\BibitemShut {NoStop}%
\bibitem [{\citenamefont {Saha}\ \emph {et~al.}(2019)\citenamefont {Saha}, \citenamefont {Wu}, \citenamefont {Su},\ and\ \citenamefont {Wang}}]{Saha2019}%
  \BibitemOpen
  \bibfield  {author} {\bibinfo {author} {\bibfnamefont {R.}~\bibnamefont {Saha}}, \bibinfo {author} {\bibfnamefont {K.}~\bibnamefont {Wu}}, \bibinfo {author} {\bibfnamefont {D.}~\bibnamefont {Su}},\ and\ \bibinfo {author} {\bibfnamefont {J.-P.}\ \bibnamefont {Wang}},\ }\bibfield  {title} {\bibinfo {title} {Tunable magnetic skyrmions in spintronic nanostructures for cellular-level magnetic neurostimulation},\ }\href@noop {} {\bibfield  {journal} {\bibinfo  {journal} {J. Phys. D: Appl. Phys.}\ }\textbf {\bibinfo {volume} {52}},\ \bibinfo {pages} {465002} (\bibinfo {year} {2019})}\BibitemShut {NoStop}%
\bibitem [{\citenamefont {Vansteenkiste}\ \emph {et~al.}(2014)\citenamefont {Vansteenkiste}, \citenamefont {Leliaert}, \citenamefont {Dvornik}, \citenamefont {Helsen}, \citenamefont {Garcia-Sanchez},\ and\ \citenamefont {Waeyenberge}}]{Mumax3}%
  \BibitemOpen
  \bibfield  {author} {\bibinfo {author} {\bibfnamefont {A.}~\bibnamefont {Vansteenkiste}}, \bibinfo {author} {\bibfnamefont {J.}~\bibnamefont {Leliaert}}, \bibinfo {author} {\bibfnamefont {M.}~\bibnamefont {Dvornik}}, \bibinfo {author} {\bibfnamefont {M.}~\bibnamefont {Helsen}}, \bibinfo {author} {\bibfnamefont {F.}~\bibnamefont {Garcia-Sanchez}},\ and\ \bibinfo {author} {\bibfnamefont {B.~V.}\ \bibnamefont {Waeyenberge}},\ }\bibfield  {title} {\bibinfo {title} {The design and verification of mumax3},\ }\href {https://doi.org/10.1063/1.4899186} {\bibfield  {journal} {\bibinfo  {journal} {AIP Adv.}\ }\textbf {\bibinfo {volume} {4}} (\bibinfo {year} {2014})}\BibitemShut {NoStop}%
\bibitem [{\citenamefont {Tomasello}\ \emph {et~al.}(2014{\natexlab{b}})\citenamefont {Tomasello}, \citenamefont {Martinez}, \citenamefont {Zivieri}, \citenamefont {Torres}, \citenamefont {Carpentieri},\ and\ \citenamefont {Finocchio}}]{Tomasello2014}%
  \BibitemOpen
  \bibfield  {author} {\bibinfo {author} {\bibfnamefont {R.}~\bibnamefont {Tomasello}}, \bibinfo {author} {\bibfnamefont {E.}~\bibnamefont {Martinez}}, \bibinfo {author} {\bibfnamefont {R.}~\bibnamefont {Zivieri}}, \bibinfo {author} {\bibfnamefont {L.}~\bibnamefont {Torres}}, \bibinfo {author} {\bibfnamefont {M.}~\bibnamefont {Carpentieri}},\ and\ \bibinfo {author} {\bibfnamefont {G.}~\bibnamefont {Finocchio}},\ }\bibfield  {title} {\bibinfo {title} {A strategy for the design of skyrmion racetrack memories},\ }\bibfield  {journal} {\bibinfo  {journal} {Scientific Reports}\ }\textbf {\bibinfo {volume} {4}},\ \href {https://doi.org/10.1038/srep06784} {10.1038/srep06784} (\bibinfo {year} {2014}{\natexlab{b}})\BibitemShut {NoStop}%
\bibitem [{\citenamefont {B\"{u}ttner}\ \emph {et~al.}(2017)\citenamefont {B\"{u}ttner}, \citenamefont {Lemesh}, \citenamefont {Schneider}, \citenamefont {Pfau}, \citenamefont {G\"{u}nther}, \citenamefont {Hessing}, \citenamefont {Geilhufe}, \citenamefont {Caretta}, \citenamefont {Engel}, \citenamefont {Kr\"{u}ger}, \citenamefont {Viefhaus}, \citenamefont {Eisebitt},\ and\ \citenamefont {Beach}}]{Bttner2017}%
  \BibitemOpen
  \bibfield  {author} {\bibinfo {author} {\bibfnamefont {F.}~\bibnamefont {B\"{u}ttner}}, \bibinfo {author} {\bibfnamefont {I.}~\bibnamefont {Lemesh}}, \bibinfo {author} {\bibfnamefont {M.}~\bibnamefont {Schneider}}, \bibinfo {author} {\bibfnamefont {B.}~\bibnamefont {Pfau}}, \bibinfo {author} {\bibfnamefont {C.~M.}\ \bibnamefont {G\"{u}nther}}, \bibinfo {author} {\bibfnamefont {P.}~\bibnamefont {Hessing}}, \bibinfo {author} {\bibfnamefont {J.}~\bibnamefont {Geilhufe}}, \bibinfo {author} {\bibfnamefont {L.}~\bibnamefont {Caretta}}, \bibinfo {author} {\bibfnamefont {D.}~\bibnamefont {Engel}}, \bibinfo {author} {\bibfnamefont {B.}~\bibnamefont {Kr\"{u}ger}}, \bibinfo {author} {\bibfnamefont {J.}~\bibnamefont {Viefhaus}}, \bibinfo {author} {\bibfnamefont {S.}~\bibnamefont {Eisebitt}},\ and\ \bibinfo {author} {\bibfnamefont {G.~S.~D.}\ \bibnamefont {Beach}},\ }\bibfield  {title} {\bibinfo {title} {Field-free deterministic ultrafast creation of magnetic skyrmions by spin–orbit torques},\ }\href
  {https://doi.org/10.1038/nnano.2017.178} {\bibfield  {journal} {\bibinfo  {journal} {Nature Nanotechnology}\ }\textbf {\bibinfo {volume} {12}},\ \bibinfo {pages} {1040–1044} (\bibinfo {year} {2017})}\BibitemShut {NoStop}%
\bibitem [{\citenamefont {Zhang}\ \emph {et~al.}(2015)\citenamefont {Zhang}, \citenamefont {Han}, \citenamefont {Jiang}, \citenamefont {Yang},\ and\ \citenamefont {S.~P.~Parkin}}]{Zhang2015}%
  \BibitemOpen
  \bibfield  {author} {\bibinfo {author} {\bibfnamefont {W.}~\bibnamefont {Zhang}}, \bibinfo {author} {\bibfnamefont {W.}~\bibnamefont {Han}}, \bibinfo {author} {\bibfnamefont {X.}~\bibnamefont {Jiang}}, \bibinfo {author} {\bibfnamefont {S.-H.}\ \bibnamefont {Yang}},\ and\ \bibinfo {author} {\bibfnamefont {S.}~\bibnamefont {S.~P.~Parkin}},\ }\bibfield  {title} {\bibinfo {title} {Role of transparency of platinum–ferromagnet interfaces in determining the intrinsic magnitude of the spin hall effect},\ }\href {https://doi.org/10.1038/nphys3304} {\bibfield  {journal} {\bibinfo  {journal} {Nature Physics}\ }\textbf {\bibinfo {volume} {11}},\ \bibinfo {pages} {496–502} (\bibinfo {year} {2015})}\BibitemShut {NoStop}%
\bibitem [{\citenamefont {Thiele}(1973)}]{Thiele}%
  \BibitemOpen
  \bibfield  {author} {\bibinfo {author} {\bibfnamefont {A.~A.}\ \bibnamefont {Thiele}},\ }\bibfield  {title} {\bibinfo {title} {Steady-state motion of magnetic domains},\ }\href@noop {} {\bibfield  {journal} {\bibinfo  {journal} {Phys. Rev. Lett.}\ }\textbf {\bibinfo {volume} {30}},\ \bibinfo {pages} {230} (\bibinfo {year} {1973})}\BibitemShut {NoStop}%
\bibitem [{\citenamefont {B\"uttner}\ \emph {et~al.}(2015)\citenamefont {B\"uttner}, \citenamefont {Moutafis}, \citenamefont {Schneider}, \citenamefont {Kr\"uger}, \citenamefont {G\"unther}, \citenamefont {Geilhufe}, \citenamefont {v.~Korff~Schmising}, \citenamefont {Mohanty}, \citenamefont {Pfau}, \citenamefont {Schaffert}, \citenamefont {Bisig}, \citenamefont {Foerster}, \citenamefont {Schulz}, \citenamefont {Vaz}, \citenamefont {Franken}, \citenamefont {Swagten}, \citenamefont {Kl\"aui},\ and\ \citenamefont {Eisebitt}}]{Butner-Nat}%
  \BibitemOpen
  \bibfield  {author} {\bibinfo {author} {\bibfnamefont {F.}~\bibnamefont {B\"uttner}}, \bibinfo {author} {\bibfnamefont {C.}~\bibnamefont {Moutafis}}, \bibinfo {author} {\bibfnamefont {M.}~\bibnamefont {Schneider}}, \bibinfo {author} {\bibfnamefont {B.}~\bibnamefont {Kr\"uger}}, \bibinfo {author} {\bibfnamefont {C.~M.}\ \bibnamefont {G\"unther}}, \bibinfo {author} {\bibfnamefont {J.}~\bibnamefont {Geilhufe}}, \bibinfo {author} {\bibfnamefont {C.}~\bibnamefont {v.~Korff~Schmising}}, \bibinfo {author} {\bibfnamefont {J.}~\bibnamefont {Mohanty}}, \bibinfo {author} {\bibfnamefont {B.}~\bibnamefont {Pfau}}, \bibinfo {author} {\bibfnamefont {S.}~\bibnamefont {Schaffert}}, \bibinfo {author} {\bibfnamefont {A.}~\bibnamefont {Bisig}}, \bibinfo {author} {\bibfnamefont {M.}~\bibnamefont {Foerster}}, \bibinfo {author} {\bibfnamefont {T.}~\bibnamefont {Schulz}}, \bibinfo {author} {\bibfnamefont {C.~A.~F.}\ \bibnamefont {Vaz}}, \bibinfo {author} {\bibfnamefont {J.~H.}\ \bibnamefont {Franken}}, \bibinfo {author}
  {\bibfnamefont {H.~J.~M.}\ \bibnamefont {Swagten}}, \bibinfo {author} {\bibfnamefont {M.}~\bibnamefont {Kl\"aui}},\ and\ \bibinfo {author} {\bibfnamefont {S.}~\bibnamefont {Eisebitt}},\ }\bibfield  {title} {\bibinfo {title} {Dynamics and inertia of skyrmionic spin structures},\ }\href@noop {} {\bibfield  {journal} {\bibinfo  {journal} {Nat. Phys.}\ }\textbf {\bibinfo {volume} {11}},\ \bibinfo {pages} {225} (\bibinfo {year} {2015})}\BibitemShut {NoStop}%
\bibitem [{\citenamefont {Mukai}\ and\ \citenamefont {Leonov}(2024)}]{Mukai2024}%
  \BibitemOpen
  \bibfield  {author} {\bibinfo {author} {\bibfnamefont {N.}~\bibnamefont {Mukai}}\ and\ \bibinfo {author} {\bibfnamefont {A.~O.}\ \bibnamefont {Leonov}},\ }\bibfield  {title} {\bibinfo {title} {“polymerization” of bimerons in quasi-two-dimensional chiral magnets with easy-plane anisotropy},\ }\href {https://www.mdpi.com/2079-4991/14/6/504} {\bibfield  {journal} {\bibinfo  {journal} {Nanomaterials}\ }\textbf {\bibinfo {volume} {14}} (\bibinfo {year} {2024})}\BibitemShut {NoStop}%
\bibitem [{com(2023)}]{comsol2023}%
  \BibitemOpen
  \href@noop {} {\bibinfo {title} {Comsol modeling software}},\ \bibinfo {howpublished} {\url{https://www.comsol.com}} (\bibinfo {year} {2023}),\ \bibinfo {note} {cOMSOL, Inc. Retrieved (2023)}\BibitemShut {NoStop}%
\end{thebibliography}%

\newpage

\end{document}